\documentclass[preprint,preprintnumbers,amsmath,amssymb]{revtex4}
\usepackage{graphicx}
\usepackage{dcolumn}
\usepackage{bm}
\usepackage{color}
\usepackage{bm}
\usepackage[active]{srcltx}

\def\|{|\!|}

\newcommand{\ba}{\begin{array}}\newcommand{\ea}{\end{array}}
\newcommand{\be}{\begin{equation}}\newcommand{\ee}{\end{equation}}
\newcommand{\bea}{\begin{eqnarray}}\newcommand{\eea}{\end{eqnarray}}
\newcommand{\brr}{\begin{array}}\newcommand{\err}{\end{array}}
\newcommand{\bit}{\begin{itemize}}\newcommand{\eit}{\end{itemize}}
\newcommand{\ben}{\begin{enumerate}}\newcommand{\een}{\end{enumerate}}

\def\beq{\begin{equation}}
\def\eeq{\end{equation}}

\definecolor{darkred}{rgb}{.8,0,0}

\begin{document}

\title{R\'{e}nyi's information transfer between financial time series}

\author{Petr~Jizba}
\email{jizba@physik.fu-berlin.de}
\affiliation{ITP, Freie Universit\"{a}t Berlin, Arnimallee 14
D-14195 Berlin, Germany\\} \affiliation{FNSPE, Czech Technical
University in Prague, B\v{r}ehov\'{a} 7, 115 19 Praha 1, Czech
Republic}

\author{Hagen~Kleinert}
\email{kleinert@physik.fu-berlin.de}
\affiliation{ITP, Freie Universit\"{a}t Berlin, Arnimallee 14
D-14195 Berlin, Germany \\} \affiliation{ICRANeT, Piazzale della
Republica 1, 10 -65122, Pescara, Italy}
\author{Mohammad~Shefaat}
\email{mohammad.shefaat@quirinbank.de}
\affiliation{Quirin Bank AG, Kurf\"{ur}stendamm 119, 10711 Berlin,
Germany\\}

\begin{abstract} \vspace{3mm} In this paper, we quantify the statistical coherence between
financial time series by means of the R\'{e}nyi entropy. With the help of Campbell's
coding theorem we show that the R\'{e}nyi entropy selectively emphasizes only certain sectors of the underlying empirical distribution while strongly suppressing others. This accentuation is controlled with R\'{e}nyi's parameter $q$. To tackle the issue of the information flow between time series we formulate the concept of R\'{e}nyi's transfer entropy as a measure of information that is transferred only between certain parts of underlying distributions.
This is particularly pertinent in financial time series where the knowledge of
marginal events such as spikes or sudden jumps is of a crucial importance.
We apply the R\'{e}nyian information flow to stock market time series from $11$ world stock indices as sampled at a daily rate  in the time period 02.01.1990 - 31.12.2009.  Corresponding {\em heat maps} and {\em net information flows} are represented graphically. A detailed discussion of the transfer entropy between the DAX and S\&P500 indices based on minute tick data gathered in the period from 02.04.2008 to 11.09.2009 is also provided. Our analysis shows that the bivariate information flow between world markets is strongly asymmetric with a distinct information surplus  flowing from the Asia--Pacific region to both European and US markets. An important yet less dramatic excess of information also flows from Europe to the US. This is particularly clearly seen from a careful analysis of R\'{e}nyi information flow between the DAX and S\&P500 indices.



\vspace{4mm}


\end{abstract}

\maketitle

\section{Introduction~\label{sec1}}

The evolution of many complex systems in natural,
economical, and social sciences is usually presented in the form
of time series. In order to analyze time series, several of statistical measures have been introduced in the literature. These include such concepts as probability distributions~\cite{1,2},
autocorrelations~\cite{2}, multi-fractals~\cite{3}, complexity~\cite{4,5}, or entropy densities~\cite{5}. Recently, it has been pointed out that the {\em transfer entropy} (TE) is a very useful instrument in quantifying statistical coherence between time evolving statistical systems~\cite{schreiber,marschinski,kwon2}. In particular, in Schreiber's
paper~\cite{schreiber} it was demonstrated  that TE is especially expedient when global properties of time series are analyzed. Prominent applications are in multivariate
analysis of time series, including e.g., study of multichannel physiological data or bivariate analysis of historical stock exchange indices. Methods based on TE have substantial computational advantages which are particularly important in analyzing a large amount of data. In all past works, including~\cite{vejmelka,kwon2,lungarella}, the emphasis has been on various generalizations of transfer entropies that were firmly rooted in the framework of Shannon's information theory. These, so called Shannonian transfer entropies
are, indeed, natural candidates due to their ability to quantify in a non-parametric and in
explicitly non-symmetric way the flow of information between two time series. An ideal
testing ground for various TE concepts are financial-market time series because of the
immense amount of electronically recorded financial data.

Recently, economy has become an active research area for physicists. They have investigated
stock markets using statistical-physics methods, such as the percolation theory,
multifractals, spin-glass models, information theory, complex networks, path integrals,
etc.. In this connection the name econophysics has been coined to denote this new hybrid
field on the border between statistical physic and (quantitative) finance.
In the framework of econophysics it has became steadily evident that the market interactions are highly nonlinear, unstable, and long-ranged. It has also became apparent that all agents (e.g., companies) involved in a given stock market exhibit interconnectedness and correlations which represent important internal force of the market. Typically one uses correlation functions to study the internal cross-correlations between various market activities. The correlation functions have, however, at least two limitations: First, they measure only linear relations, although it is clear that linear models do not faithfully reflect real market interactions. Second, all they determine is whether two time series (e.g., two stock-index series) have correlated movement. They, however, do not indicate
which series affects which, or in other words, they do not provide any directional information about cause and effect. Some authors use such concepts as time-delayed correlation or time-delayed mutual information in order to construct asymmetric ``correlation'' matrices with inherent directionality. This procedure is in many respects {\em ad hoc} as it does not provide any natural measure (or quantifier) of the information flow between involved series.

In the present paper we study multivariate properties of stock-index time with the help
of econophysics paradigm. In order to quantify the information flow between two or more
stock indices we generalize Schreibers' Shannonian transfer entropy to  R\'{e}nyi's
information setting. With this we demonstrate that the corresponding new transfer entropy
provides more detailed information concerning the excess (or lack) of information
in various parts of the underlying distribution resulting from updating the distribution on the condition that a second time series is known. This is particularly relevant in the
context of financial time series where the knowledge of tale-part (or marginal) events such as spikes or sudden jumps bears direct implications, e.g., in various risk-reducing formulas in portfolio theory.

The paper is organized as follows: In Section~\ref{sec2} we provide some information-theoretic
background on Shannon and R\'{e}nyi entropies (RE's). In particular, we identify the
{\em conditional} R\'{e}nyi's entropy with the information measure introduced in Ref.~\cite{JA}.
Apart from satisfying the chain rule (i.e., rule of additivity of information) the latter has many desirable properties that are to be expected from a
conditional information measure. Another key concept --- the {\em mutual} R\'{e}nyi entropy, is then
introduced in a close analogy with Shannon's case. The ensuing properties are
also discussed. Shannonian {\em transfer} entropy  of Schreiber is briefly reviewed in
Section~\ref{sec4}. There we also comment on effective transfer entropy of  Marchinski
{\em et all}. The core quantity of this work --- the R\'{e}nyian {\em transfer} entropy
(RTE), is motivated and derived in Section~\ref{sec7}. In contrast to Shannonian case,
the R\'{e}nyian transfer entropy  is generally not positive semi-definite. This is because
RE non-linearly emphasizes different parts of involved probability density functions
(PDF's). With the help of Campbell's coding theorem we show that the RTE rates a gain/loss in risk involved in a next-time-step behavior in a given stochastic process, say $X$,
resulting from learning a new information, namely historical behavior of another (generally
cross-correlated) process, say $Y$. In this view the RTE can serve as a convenient
{\em rating} factor of a {\em riskiness} in inter-connected markets. We also show that R\'{e}nyian transfer
entropy allows to amend spurious effects caused by a finite size of a real
data set which in Shannon's context must be, otherwise, solved by means of the surrogate
data technique and ensuing effective transfer entropy. In Section~\ref{sec10} we demonstrate the
usefulness and formal consistency of RTE by analyzing
cross-correlations in various international stock markets. On a qualitative level we
use $183.308$ simultaneously recorded data points of the eleven stock exchange indices,
sampled at a daily (end-of-trading day) rate to construct the {\em heat maps} and
{\em net flows} for both Shannon's and R\'{e}nyi's information flows. On a quantitative level
we explicitly discuss  time series from the DAX and  S\&P500 market indices gathered on a minute-tick
basis in the period from December 1990 till November 2009 in the German stock exchange market
(Deutsche B\"{o}rse). Presented calculations of R\'{e}nyi and Shannon
transfer entropies are based on symbolic coding computation with the open source software $R$.
Our numerical results imply that RTE's among world markets are
typically very asymmetric. For instance, we show that
there is a strong surplus of an information flow from the Asia-Pacific region to both Europe and the
U.S. A surplus of the information flow can be also observed to exists from Europe to the U.S.
In this last case the substantial volume of transferred information comes from tail-part
(i.e., risky part) of  underlying asset distributions. So, despite the fact that the U.S.
contributes more than half of the world trading volume, this is not so with information flow.

Further salient issues, such as dependence of RTE on R\'{e}nyi's $q$
parameter or on the data block length are numerically also investigated.
In this connection we find that the cross-correlation between DAX and S\&P500 has a long-time memory which is around 200-300 mins. This should be contrasted with typical memory of stock returns
which are of the order of seconds or maximally few minutes. Various remarks and generalizations are proposed in the concluding Section~\ref{SEc12}.
For reader's convenience we give in Appendix~A a brief dictionary of market Indices
used in the main text and in Appendix~B we tabulate an explicit values of effective transfer entropies
used in the construction of heat maps and net information flows.

\section{Information-theoretic entropies of Shannon and R\'{e}nyi ~\label{sec2}}

In order to express numerically  an amount of information that is shared or transferred
between various data sets (e.g., two or more random processes), one commonly resorts to
information theory and especially to the concept of entropy. In this section we briefly
review some essentials of Shannon's and R\'{e}nyi's entropy that will be needed in
following sections.

\subsection{Shannon's entropy~\label{sec2b}}

The entropy concept was originally introduced by Clausius~\cite{huang} in the
framework of thermodynamics. By analyzing a Carnot engine he was able to identify a new state
function which never decreases in isolated systems. The microphysical origin of
Clausius' phenomenological entropy was clarified more than $20$ years later in works of Boltzman
and (yet later) Gibbs who associated Clausius entropy with the number of allowed
microscopic states compatible with a given observed macrostate. The ensuing {\em Boltzmann--Gibbs
entropy} reads
\begin{equation} H_{BG}({\mathcal{P}}) \ = \ -k_B \sum_{x\ \! \in \ \! {X}}^{W}p(x)\,\ln p(x) \, ,
\label{II.1.b}
 \\
\end{equation}
where $k_B$ is Boltzmann's constant, $X$ is the set of all accessible microstates compatible
with whatever macroscopic observable (state variable) one controls
and $W$ denotes the number of microstates.

It should be said that the passage from Boltzmann--Gibbs to Clausius entropy is established only
when the conditional extremum ${\mathcal{P}}_{\rm ex}$ of $H_{BG}$ subject to the
constraints imposed by observed state variables is inserted back into $H_{BG}$. Only when this
{\em maximal entropy prescription}~\cite{gibbs} is utilized $H_{BG}$ turns out to
be a thermodynamic state function and not mere functional on a probability space.

In information theory, on the other hand, the interest was in an optimal coding of a given
source data. By {\em optimal code} is meant the shortest averaged code from which one
can uniquely decode the source data. Optimality of coding was solved by Shannon in his 1948
seminal paper~\cite{shannon}. According to Shannon's {\em source coding
theorem}~\cite{shannon,SW}, the quantity
\begin{equation} H({\mathcal{P}}) \ = \ -\sum_{x\ \! \in \ \! {X}}^{W}p(x)\,\log_2 p(x) \, ,
\label{II.2.a}
 \\
\end{equation}
corresponds to the averaged number of bits needed to optimally encode (or ``zip") the source
dataset $X$ with the source probability distribution
${\mathcal{P}}(X)$. On a quantitative level (\ref{II.2.a}) represents  (in bits) the minimal
number of binary (yes/no) questions that brings us from our present state of
knowledge about the system $X$ to the one of certainty~\cite{shannon,renyi1,ash}. It should
be stressed that in Shannon's formulation $X$ represents a discrete set (e.g.,
processes with discrete time), and this will be also the case here. Apart from the foregoing
{\em operational} definitions, Eq.~(\ref{II.2.a}) has also several axiomatic underpinnings.
Axiomatic approaches were advanced by Shannon~\cite{shannon,SW}, Khinchin~\cite{khinchin},
Fadeev~\cite{fadeev} an others~\cite{others}. The quantity (\ref{II.2.a}) has became
known as Shannon's entropy (SE).

There is an intimate connection between Boltzmann--Gibbs entropy and Shannon's entropy.
In fact, thermodynamics can be viewed as a specific application of Shannon's information
theory: the thermodynamic entropy may be interpreted (when rescaled to ``bit'' units) as
the amount of Shannon information needed to define the detailed microscopic state of the
system, that remains ``uncommunicated" by a description that is solely in terms of
thermodynamic state variables~\cite{szilard,brillouin,jaynes}.

Among important properties of SE is its concavity in $\mathcal{P}$, i.e. for any pair of
distributions $\mathcal{P}$ and $\mathcal{Q}$, and a real number $0\leq \lambda\leq 1$
holds
\begin{equation} H(\lambda \mathcal{P} + (1-\lambda)\mathcal{Q})\ \geq \ \lambda H(\mathcal{P} ) + (1-\lambda)H(\mathcal{Q})\, .
\label{I.19.a} \end{equation}
Eq.~(\ref{I.19.a}) follows from Jensen's inequality and a convexity of $x\log x$ for $x> 0$.
Concavity is an important concept since it ensures that any maximizer found by the
methods of the differential calculus yields an absolute maximum rather than a relative
maximum or minimum or saddle point. At the same time it is just a sufficient (i.e., not
necessary) condition guarantying a unique maximizer. It is often customary to denote
SE of the source $X$ as $H(X)$ rather than $H(\mathcal{P})$. Note that SE is generally not
convex in $X$!

It should be stressed that the entropy (\ref{II.2.a}) really represents a self-information:
the information yielded by a random process about itself. A step further from a
self-information offers the {\em joint entropy} of two random variables $X$ and $Y$ which
is defined as
\begin{equation}
H({X}\cap {Y}) \ = \ -\!\!\!\!\sum_{x\ \!\in \ \!{X}, \
\!y\ \!\in \ \!{Y}} p(x,y) \log_2{p(x,y)}\, , \label{eq:jointEntropy}
\end{equation}
and which represents the amount of information gained by observing jointly two (generally
dependent or correlated) statistical events.

A further concept that will be needed here is the
{\em conditional entropy} of $X$ given $Y$, which can be
motivated as follows: Let us have two statistical events $X$ and $Y$ and let event
$Y$ has a sharp value $y$, then the gain of information obtained by observing $X$ is
\begin{eqnarray}
H(X|Y=y) \ = \ -\!\!\!\!\sum_{x\ \!\in \ \!{X}} p(x|y) \log_2{p(x|y)}\, .
\end{eqnarray}
Here the conditional probability $p(x|y) = p(x,y)/p(y)$. For general random $Y$ one defines
the conditional entropy as the averaged Shannon entropy yielded by
$X$ under the assumption that the value of $Y$ is known, i.e.
%
\begin{eqnarray}
H(X|Y) \ = \ \sum_{y\
\!\in \ \!{Y}}p(y)H(X|Y=y) \ = \ -\!\!\!\!\sum_{x\ \!\in \ \!{X}, \ \!y\ \!\in \ \!{Y}} p(x,y) \log_2{p(x|y)}\, .
\label{II.A.5a}
\end{eqnarray}
From (\ref{II.A.5a}), in particular, follows that
\begin{eqnarray}
H({X}\cap {Y}) \ = \  H({Y}) \ + \ H({X}|{Y}) \ = \ H({X}) \ + \ H({Y}|{X}) \, .
\label{II.2.6.a}
\end{eqnarray}
Identity (\ref{II.2.6.a}) is known as additivity (or chain) rule for Shannon's entropy.
In statistical thermodynamics this rule allows to explain, e.g., Gibbs paradox. Applying
Eq.~(\ref{II.2.6.a}) iteratively, we obtain:
\begin{eqnarray}
H({X}_1\cap {X}_2 \cap \cdots \cap X_n) \ &=& \
H({X}_1) \ + \ H({X}_2|X_1) \ + \ H({X}_3|X_1\cap {X}_2) \ + \
\cdots \nonumber \\ &=& \ \sum_{i}^n H(X_i|X_1\cap {X}_2 \cap \cdots \cap X_{i-1})\, .
\label{II.A.8a}
\end{eqnarray}

Another relevant quantity that will be needed is the {\em mutual information} between $X$ and $Y$.
This is defined as:
\begin{eqnarray}
I(X;Y) \ = \ \sum_{x\ \!\in \ \!{X}, \
\!y\ \!\in \ \!{Y}} p(x,y) \log_2{\frac{p(x,y)}{p(x)q(y)}}\, ,
\label{eq:mi}
\end{eqnarray}
and can be equivalently written as
\begin{eqnarray}
I(X;Y) \ = \ H(X)\ - \ H(X|Y)
\ = \ H(Y) \ - \ H(Y|X)\, .
\label{eq:mi2}
\end{eqnarray}
This shows that the mutual information measures the average reduction in
uncertainty (i.e., gain in information) about
$X$ resulting from observation of $Y$. Of course, the amount of information contained
in $X$ about itself is just the Shannon entropy:
\begin{eqnarray}
I(X;X) \ = \ H(X)\,.
\end{eqnarray}

Notice also that from Eq.~(\ref{eq:mi}) follows  $I(X;Y)= I(Y;X)$ and so $X$
provides the same amount of information on $Y$ as $Y$ does on $X$.
For this reasons the mutual
information is not a useful measure to quantify a flow of information.
In fact, the flow of information should be by its very definition directional.

In the following we will also find useful the concept of {\em conditional mutual entropy}
between $X$ and $Y$ given $Z$ which is defined as
\begin{eqnarray}
I(X;Y|Z)\ &=& \ H(X|Z)\ - \
H(X|Y\cap Z)\, ,\nonumber \\ &=& \ I(X;Y\cap Z) \ - \ I(X;Y) \, .
\label{eq:condMII}
\end{eqnarray}
The latter quantifies the averaged mutual information between $X$ and $Y$
provided that $Z$ is known. Applying (\ref{eq:condMII}) and (\ref{eq:mi2})
iteratively we may write
\begin{eqnarray}
I(X;Y_1\cap \cdots \cap Y_n|Z_1\cap \cdots \cap Z_m)\ &=& \
H(X|Z_1\cap \cdots \cap Z_m)\nonumber \\ &-& \ H(X|Y_1\cap
\cdots \cap Y_n\cap Z_1\cap \cdots \cap Z_m)\nonumber \\[2mm]
&=& \ I(X;Y_1\cap \cdots \cap Y_n \cap Z_1\cap \cdots
\cap Z_m)\nonumber \\ &-&  \ I(X;Z_1\cap \cdots \cap Z_m) \, .
\label{eq:condMIIII}
\end{eqnarray}

For further details on the basic concepts of Shannon's information theory, we refer
the reader to classical books, e.g., Ash~\cite{ash} and, more recently, Csisz\'{a}r and
Shields~\cite{C-S}.

\subsection{R\'{e}nyi's entropy~\label{sec2c}}

\begin{figure}[ht]
 \begin{center}
\includegraphics*[width=10cm]{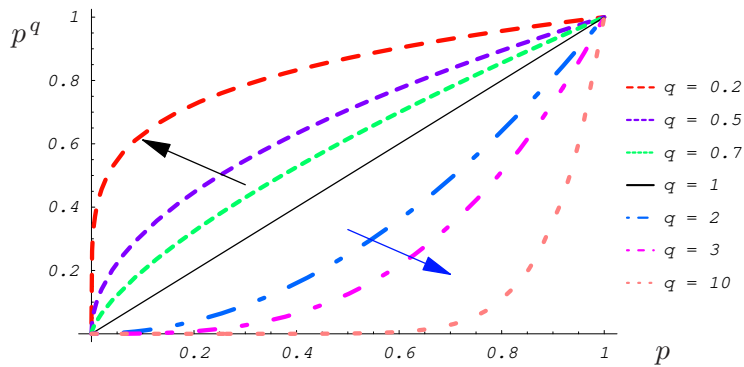}
\end{center} \vspace{-.5cm}
\caption{The function $p^q$ for event probability $p$ and varying R\'{e}nyi's parameter $q$. Arrows indicate
decreasing values of $q$ for $0<q<1$ (dark arrow) or increasing values of $q$ for $q>1$ (lighter arrow).}
\label{fig2}
\begin{picture}(20,7)
\put(105,65){ $p$ }
\put(-135,185){$p^{\,q}$}
\end{picture}
\end{figure}

\begin{figure}[ht]
\begin{center} \includegraphics*[width=8cm]{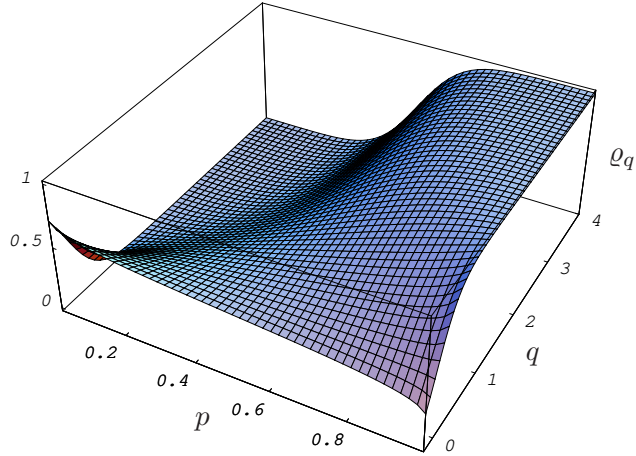}
\end{center} \vspace{-.5cm}
\caption{A plot of the escort distribution for two-dimensional ${\mathcal{P}}$: $\varrho_q =
p^{\, q}/(p^{\, q} + (1-p)^q)$. } \label{fig1}
\begin{picture}(20,7) \put(90,80){ $q$ }
\put(-35,55){ $p$ } \put(125,155){$\varrho_q$}
\end{picture}
\end{figure}

R\'{e}nyi introduced in Refs.~\cite{renyi,renyi0} a one-parameter family of information
measures presently known as {\em R\'{e}nyi entropies}~\cite{renyi,JA}.  In practice,
however, only a singular name --- R\'{e}nyi's entropy --- is used. RE of
order $q$ $(q > 0)$ of a distribution ${\mathcal{P}}$ on a finite set ${X}$  is defined as
\begin{equation}
S_q^{(R)}({\mathcal{P}})\ = \ \frac{1}{1-q}\log_2 \sum_{x\ \! \in \ \! {X}} p^{\,q}(x)\, .
\label{renyi}
\end{equation}
For RE (\ref{renyi}) one can also formulate source coding theorem. While in the Shannon
case the cost of a code-word is a linear function of the length --- so the optimal code
has a minimal cost out of all codes,
in the R\'{e}nyi case the cost of a code-word is an exponential function of its
length~\cite{campbell,aczel,bercher}. This is, in a nutshell, an essence of the so-called
Campbell's coding theorem (CCT). According to this RE corresponds to the averaged
number of bits needed to optimally encode the discrete source $X$ with the probability
${\mathcal{P}}(X)$, provided that the codeword-lengths are exponentially
weighted~\cite{expweight}. From the form (\ref{renyi}) one can easily see
that for $q>1$ RE depends more on the probabilities of the more probable values and
less on the improbable ones. This dependence is more pronounced for higher $q$.
On the other hand, for $0<q<1$ marginal
events are accentuated with decreasing $q$.  In this connection we should also
point out  that Campbell's coding theorem for RE is equivalent to Shannon's coding theorem for SE
provided one uses instead of $p(x)$ the {\em escort distribution}~\cite{bercher}:
\begin{eqnarray}
\varrho_q(x) \ \equiv \ \frac{p^{\,q}(x)}{\sum_{x\ \! \in \ \! {X}}
p^{\,q}(x)}\, .
\label{II.15.a}
\end{eqnarray}
The PDF $\varrho_q(x)$ was first introduced by R\'{e}nyi~\cite{renyi0} and in the physical context brought by Beck, Schl\"{o}gl, Kadanoff and others (see, e.g., Refs.~\cite{beck,kadanov}). Note (cf. Fig.~\ref{fig2})
that for $q>1$ the escort distribution emphasizes the more probable events and de-emphasizes more improbable ones. This trend is more pronounced for higher values of $q$. For $0<q<1$ the
escort distribution accentuates more improbable (i.e., marginal or rare) events. This dependence is
more pronounced for decreasing $q$. This fact is clearly seen on Fig.~\ref{fig1}.
So by choosing different $q$ we can ``scan" or ``probe" different
parts of the involved PDF's.

It should be stressed that apart from CCT, RE has yet further operational definitions,
e.g., in the theory of guessing~\cite{arikan}, in the buffer overflow problem~\cite{jelinek} or in the theory of error block coding~\cite{csiszar}.
RE is also underpinned with various axiomatics~\cite{renyi,renyi0,daroczi}.
In particular, it satisfies identical Khinchin axioms~\cite{khinchin} as Shannon's
entropy save for the additivity axiom (chain rule)~\cite{JA,jizba2,jizba3}:
\begin{equation}
S_q^{(R)}(X\cap Y) \ = \ S_q^{(R)}(Y) \ + \  S_q^{(R)}(X|Y)\, ,
\label{III.B.15a}
\end{equation}
where the conditional entropy $S_q^{(R)}(X|Y)$ is defined with the help of
the escort distribution (\ref{II.15.a}) (see, e.g., Refs.~\cite{JA,tsallis,beck}).
For $q\to 1$ RE reduces to the Shannon entropy:
\begin{equation}
S_1^{(R)} \ = \  \lim_{q\to 1}S_q^{(R)} \ = \ H\, ,
\end{equation}
as one can easily verify with l'H{o}spital's rule.

%
We define the {\em joint R\'{e}nyi entropy} (or the {\em joint entropy} of order $q$) for two random variables ${X}$  and ${Y}$ in a natural way as:
\begin{equation}
S_q^{(R)}({X}\cap{Y}) \ = \ \frac{1}{1-q}\log_2 \sum_{x\ \! \in \ \! {X}} p^{\,q}(x,y)\, .
\label{re.jointEntropy}
\end{equation}
The {\em conditional entropy} of order $q$ of $X$ given $Y$ is similarly as in the Shannon case defined
as the averaged R\'{e}nyi's entropy yielded by $X$ under the assumption
that the value of $Y$ is known. As shown in Refs.~\cite{JA,jizba2,golshani} this has the form
\begin{eqnarray}
S_q^{(R)}(X|Y)\ &=& \ \frac{1}{1-q}\log_2 \frac{\sum_{x\ \! \in \
\! {X}, \ \! y \ \! \in \ \! {Y}}p^{\,q}(x|y) q^{\,q}(y)}{\sum_{y\ \! \in \ \! {Y}} q^{\,q}(y)}
\nonumber \\[2mm]
&=& \ \frac{1}{1-q}\log_2 \frac{\sum_{x\ \! \in \ \! {X}, \ \! y
\ \! \in \ \! {Y}}p^{\,q}(x,y)}{\sum_{y\ \! \in \ \! {Y}} q^{\,q}(y)}\, .
\label{re.condEntropy}
\end{eqnarray}
In this connection it should be mentioned that several alternative definitions of the conditional RE exist (see, e.g., Refs.~\cite{renyi0,cahin,csiszar}), but the formulation (\ref{re.condEntropy}) differs from other versions
in a few important ways that will show up to be desirable in the following considerations.
The conditional entropy defined in (\ref{re.condEntropy}) has the following important properties, namely~\cite{JA,golshani}
%
\begin{description}
 \item[--]$0 \leq S_q^{(R)}(X|Y) \leq  \log_2 n$, where $n$ is a number of elements in $X$,
  \item[--] $S_q^{(R)}(X|Y) =  0$ only when $Y$ uniquely
determines $X$ (i.e.,  no gain in information),
  \item[--] $\lim_{q\rightarrow 1} S_q^{(R)}(X|Y) = H(X|Y)$,
  \item[--] when $X$ and $Y$ are independent then $S_q^{(R)}(X|Y)  =
S_q^{(R)}(X)$ .
 \end{description}

Unlike Shannon's case one cannot, however, deduce that the equality $S_q^{(R)}(X|Y)  =  S_q^{(R)}(X)$ implies
independency between event $X$ and $Y$. Also the inequality $S_q^{(R)}(X|Y)  \leq
S_q^{(R)}(X)$ (i.e., an extra knowledge about $Y$ lessens our ignorance about $X$) does not hold here
in general~\cite{renyi0,JA}. The latter two properties may seem as a serious
flaw. We will now argue that this is not the case and, in fact, it is even desirable.

First, in order to understand why $S_q^{(R)}(X|Y)  =  S_q^{(R)}(X)$ does not imply independency
between $X$ and $Y$ we define the information-distribution function
\begin{eqnarray}
{\mathcal{F}}_{\mathcal{P}}(x) \ = \ \sum_{-\log_2p(z) < x}p(z)\, ,
\end{eqnarray}
which represents the total probability caused by events with information
content $H(z) = -\log_2p(z) < x$. With this we have
\begin{eqnarray}
2^{(1-q)x} d {\mathcal{F}}_{\mathcal{P}}(x) \ =  \sum_{x \leq H(z) < x+ dx}2^{(1-q)H(z)}  p(z) \ =  \sum_{x
\leq H(z) < x+ dx} p^{\, q}(z)\, , \end{eqnarray}
and thus
\begin{eqnarray}
S_q^{(R)}(X) \ = \ \frac{1}{1-q}\log_2\left( \int_0^{\infty} 2^{(1-q)x} d
{\mathcal{F}}_{\mathcal{P}}(x)  \right)\, .
\end{eqnarray}

Taking the inverse Laplace transform with the help of the so-called {\em Post's inversion formula}~\cite{post}
we obtain
\begin{eqnarray} {\mathcal{F}}_{\mathcal{P}}(x) \ = \
\lim_{k\rightarrow \infty} \frac{(-1)^k}{k!} \left(\frac{k}{x \ln 2}
\right)^{k+1} \left. \frac{\partial^{k}}{\partial q^{k}}
 \left[ \frac{2^{(1-q)S_q^{(R)}(X)}}{(q-1)}\right]\right|_{q \ = \ k/(x \ln 2) +1} .
\label{III.B.20a}
\end{eqnarray}
Analogous relation holds also for ${\mathcal{F}}_{\mathcal{P}|\mathcal{Q}}(x)$ and associated $S_q^{(R)}(X|Y)$. As a result we see that when working with $S_q^{(R)}$ of different orders we receive much more information on underlying distribution than when we restrict our investigation to only one $q$ (e.g., to only Shannon's entropy). In addition, Eq.~(\ref{III.B.20a}) indicates that we need all $q > 1$ (or equivalently all $0< q <1$, see~\cite{note}) in order to uniquely identify the underlying PDF.

In view of Eq.~(\ref{III.B.20a}) we see that the equality between $S_q^{(R)}(X|Y)$  and $S_q^{(R)}(X)$ at some neighborhood of $q$ merely implies that ${\mathcal{F}}_{\mathcal{P}|\mathcal{Q}}(x) = {\mathcal{F}}_{\mathcal{P}}(x)$ for some $x$. This naturally does not ensure independency between $X$ and $Y$. We need equality $S_q^{(R)}(X|Y)= S_q^{(R)}(X)$ for all $q>1$ (or for all $0< q <1$) in order to secure that ${\mathcal{F}}_{\mathcal{P}|\mathcal{Q}}(x) = {\mathcal{F}}_{\mathcal{P}}(x)$  holds for all $x$ which would in turn guarantee that ${\mathcal{P}}(X) = {\mathcal{P}}(X| Y)$. Therefore, all RE with $q>1$ (or all with $0< q <1$) are generally required  to deduce from $S_q^{(R)}(X|Y)  = S_q^{(R)}(X)$ an independency between $X$ and $Y$.

%
In order to understand the meaning of the inequality $S_q^{(R)}(X|Y)  \leq S_q^{(R)}(X)$ we first introduce the concept of mutual information. The {\em mutual information of order} $q$ between $X$ and $Y$ can be defined as (cf. Eq.~(\ref{eq:mi2}))
\begin{eqnarray}
I_q^{(R)}(X;Y) \ &=& \ S_q^{(R)}(X)\ - \
S_q^{(R)}(X|Y)\nonumber \\[1mm] &=& \ S_q^{(R)}(X)\ + \ S_q^{(R)}(Y) \ -
\ S_q^{(R)}(X\cap Y) \, ,
\label{re.mutInf}
\end{eqnarray}
which explicitly reads
\begin{eqnarray}
I_q^{(R)}(X;Y) \ &=& \  \frac{1}{1-q}\log_2 \frac{\sum_{x\ \! \in \
\! {X}, \ \! y \ \! \in \ \! {Y}}
q^{\,q}(y) p^{\,q}(x)}{\sum_{x\ \! \in \ \! {X}, \ \! y \ \! \in \ \!
{Y}}p^{\,q}(x,y)} \nonumber \\[2mm]
&=& \ \frac{1}{1-q}\log_2 \frac{\sum_{x\ \! \in \ \! {X}, \ \! y \ \! \in
\ \! {Y}} q^{\,q}(y) p^{\,q}(x)}{\sum_{x\ \! \in \ \! {X}, \ \! y \
\! \in \ \! {Y}}q^{\,q}(y)p^{\,q}(x|y)} \, .
\label{II.20a}
\end{eqnarray}
Note that we have again the symmetry relation $I_q^{(R)}(X;Y)= I_q^{(R)}(Y;X)$ as well as the
consistency condition $I_q^{(R)}(X;X) = S_q^{(R)}(X)$. So similarly as in the Shannon case,
R\'{e}nyi's mutual information formally quantifies the average reduction in uncertainty
(i.e., gain in information) about $X$ that results from learning the value of $Y$, or vice versa.

From Eq.~(\ref{re.mutInf}) we see that the inequality in question, i.e., $S_q^{(R)}(X|Y)
\leq S_q^{(R)}(X)$ implies $I_q^{(R)}(Y;X) \geq 0$. According to (\ref{II.20a}) this can
be violated only when
\begin{eqnarray}
&&\sum_{x\ \! \in \ \! {X}} p^{\,q}(x) \ > \sum_{x\ \! \in \ \! {X}} \langle
{\mathcal{P}}^{\,q}(x|Y)\rangle_q \;\;\;\;\;\; \mbox{for}
\;\;\;\; q > 1\, , \nonumber \\[1mm]
&&\sum_{x\ \! \in \ \! {X}} p^{\,q}(x) \ < \sum_{x\ \! \in \ \! {X}} \langle
{\mathcal{P}}^{\,q}(x|Y)\rangle_q \;\;\;\;\;\; \mbox{for}
\;\;\;\; 0< q < 1\, .
\label{2.B.26a}
\end{eqnarray}
Here $\langle \ldots \rangle_q$ is an average with respect to the escort distribution
$\varrho_q(y)$ (see Eq.~(\ref{II.15.a})).

\begin{figure}[ht]
 \begin{center}
\hspace{-3.7cm}{\includegraphics*[width=11.8cm]{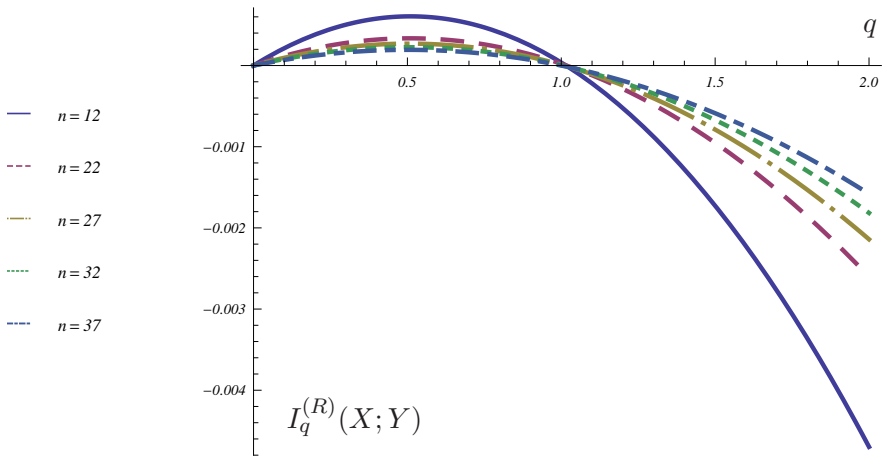}}
\end{center}
\caption{Example of a typical situation when $I_q^{(R)}(X;Y)$ is negative.  Distributions $\mathcal{P}(X)$ and $\mathcal{P}(X|Y)$ are specified in (\ref{II.20aaa}) and $\epsilon$ is chosen so that corresponding $I(X;Y) = 0$.}
\label{fig2a}
\begin{picture}(20,7)
\put(118,235){{$q$}}
\put(-100,85){{$I_q^{(R)}(X;Y)$}}
\end{picture}
\end{figure}

By taking into account properties of the escort distribution, we can deduce that
$I_q^{(R)}(X;Y)< 0$ when a larger probability events of $X$ receive by learning $Y$ a lower value.
As for the marginal events of $X$, these are by learning $Y$  indeed enhanced, but the enhancement rate
is smaller than the suppression rate of large probabilities.
%
%
%
For instance, this
happens when
\begin{eqnarray}
\mbox{\hspace{-3mm}}\mathcal{P}(X)  =  \left\{1-\epsilon,
\frac{\epsilon}{n-1}, \ldots, \frac{\epsilon}{n-1}\right\}  \mapsto \
{\mathcal{P}}(X|Y)  =  \left\{ \frac{1 - \epsilon}{2},\frac{1 - \epsilon}{2}, \frac{\epsilon}{n-2}, \ldots, \frac{\epsilon}{n-2}\right\} \! ,
\label{II.20aaa}
\end{eqnarray}
for
\vspace{-4mm}
\begin{eqnarray}
 \frac{1}{1+ \log_2(\frac{n-1}{n-2})}\ \leq\ \epsilon \ < \ 1, \,\,\,\,\,\,\,\,\,\,\,\,\,\,\,\,\, n\ > \ 2\, . \label{2.B.28a}
\end{eqnarray}
The inequality (\ref{2.B.28a}) ensures that $I(Y;X)\geq 0$ holds.  The left inequality
in (\ref{2.B.28a}) saturates when $I(Y;X) = 0$, see also Fig.~~\ref{fig2a}.
%

For $0<q<1$ is the situation analogous. Here properties of the escort distribution
imply that $I_q^{(R)}(Y;X) < 0$ when marginal events of $X$ get by learning $Y$
a higher probability.  The suppression rate for large (i.e. close-to-peak) probabilities is now smaller than the
enhancement rate of marginal events.  This happens, for example, for distributions,
\begin{eqnarray}
\mbox{\hspace{-4mm}}\mathcal{P}(X)  =  \left\{ \frac{1 - \epsilon}{2},\frac{1 - \epsilon}{2}, \frac{\epsilon}{n-2}, \ldots, \frac{\epsilon}{n-2}\right\}
\ \mapsto \
{\mathcal{P}}(X|Y)  =  \left\{1-\epsilon,
\frac{\epsilon}{n-1}, \ldots, \frac{\epsilon}{n-1}\right\}\! ,
\label{II.20aaaa}
\end{eqnarray}
with $\epsilon$ fulfilling again the inequality (\ref{2.B.28a}). This can be also directly seen from
Fig.~\ref{fig2a} when we revert the sign of $I_q^{(R)}(Y;X)$. When we set $q=1$ then both inequalities (\ref{2.B.26a}) are
simultaneously satisfied yielding $I(Y;X)=0$ --- as it should.

In contrast to a Shannonian case where the mutual information quantifies the average reduction in
uncertainty resulting from observing/learning a further information, in the
R\'{e}nyi case we should use Campbell's coding theorem in order to properly understand the meaning of
$I_q^{(R)}(Y;X)$.
According to the CCT  $S_q^{(R)}(X)$ corresponds to the minimal average cost of a
coded message with a non-linear (exponential) weighting/pricing of codeword-lengths.
While according to Shannon we never increase ignorance by learning $Y$ (i.e., possible
correlations between $X$ and $Y$ can only reduce the entropy), in R\'{e}nyi's setting
extra knowledge about $Y$ might easily increase the minimal price of coding $X$ because
of the nonlinear pricing. Since the CCT penalizes long codewords which in Shannon's coding
have low probability, the price of the $X|Y$ code may easily increase, as we have seen
in examples (\ref{II.20aaa}) and (\ref{II.20aaaa}).


In the key context of financial time series, the risk valuation of large changes such
as spikes or sudden jumps is of a crucial importance, e.g., in various risk-reducing formulas
in portfolio theory. The r\^{o}le of Campbell's pricing can in these cases be interpreted as a risk-rating
method which puts an exponential premium on rare (i.e., risky) asset
fluctuations.
%
From this point of view the mutual information $I_q^{(R)}(X;Y)$ represents a {\em rating factor}  which rates a
gain/loss in risk in $X$
resulting from learning a new information, namely information about $Y$.

The {\em conditional mutual information} of order $q$ between $X$ and $Y$ given {\em Z} is defined as
\begin{equation}
I_q^{(R)}(X;Y|Z)\ = \ S_q^{(R)}(X|Z)\ - \
S_q^{(R)}(X|Y\cap Z)\, .
\label{eq:condMI}
\end{equation}
Note that because of a validity of the chain rule (\ref{III.B.15a}), relations (\ref{II.A.8a}) and
(\ref{eq:condMIIII}) also hold true for the RE.

To close this section, we shall stress that information entropies are primarily important because
there are various coding theorems which endow them with an operational (that is,
experimental) meaning, and not because of intuitively pleasing aspects of their definitions.
While coding theorems do exist both for the Shannon entropy and the R\'{e}nyi entropy there are (as
yet) no such theorems for Tsallis', Kaniadakis', Naudts' and other currently popular entropies.
The information-theoretic significance of such entropies is thus not obvious. Since
the information-theoretic aspect of entropies is of a crucial importance here, we will in the
following focus only on the SE and the RE.

\section{Fundamentals of Shannonian transfer entropy~\label{sec4} }

\subsection{Shannonian transfer entropy~\label{sec4a}}

As seen in Section~\ref{sec2b}, the mutual information $I(X;Y)$ quantifies the decrease of
uncertainty about $X$ caused by the knowledge of $Y$. One could be thus tempted to use
it as a measure of an informational transfer in general complex systems. A major problem,
however, is that Shannon's mutual information contains no inherent directionality since
$I(X;Y) = I(Y;X)$. Some early attempts tried to resolve this complication by artificially
introducing the directionality via time-lagged random variables. In this way one may
define, for instance, the {\em time-lagged mutual } (or {\em directed Kullback--Leibler})
{\em information} as
\begin{equation}
I(X;Y)_{t,\tau} \ = \ \sum p(x_t,y_{t-\tau})
\log_2{\frac{p(x_t,x_{t-\tau})}{p(x_t)q(y_t)}}\, .
\label{III.A.1.a}
\end{equation}
The later describes the average gain of information when replacing the product probability
${\mathcal{P}}_{t}\times{\mathcal{Q}}_{t}  = \{ p(x_t)q(y_t); \;
x_t \in X_t, y_{t}\in Y_{t}\}$ by the joint probability
${\mathcal{P}}_t\cap {\mathcal{Q}}_{t-\tau} =
\{p(x_t,y_{t-\tau});\; x_t \in X_t, y_{t-\tau}\in Y_{t-\tau}\}$.
So the information gained is due to cross-correlation effect between
random variables $X_t$ and $Y_t$ (respectively,
$Y_{t-\tau}$). It was, however, pointed out in Ref.~\cite{schreiber}
that prescriptions such as (\ref{III.A.1.a}), though directional, also take
into account  some part of
the information that is statically shared between the two random processes $X$
and $Y$. In other words, these prescriptions do not produce statistical
dependences that truly
originate only in the stochastic random process $Y$, but they do include
the effects of a common history (such as, for example,  in the case of a common
external driving force).

For this reason, Schreiber introduced in Ref.~\cite{schreiber} the concept of
(Shannonian) transfer entropy (STE). The latter, apart from directionality, accounts only for the
cross-correlations between statistical time series $X$ and $Y$ whose genuine
origin is in the ``source" process $Y$. The essence of the approach is the following. Let us have two
time sequences described by stochastic random variables $X_t$ and $Y_t$. Let
us assume further that the time steps (data ticks) are discrete with the size of an elementary time lag $\tau$
and with $t_n = t_0 + n\tau$ ($t_0$ is a reference time).

The transfer entropy $T_{Y \rightarrow X}(m,l)$ can then be defined as
\begin{eqnarray}
\mbox{\hspace{-3mm}}T_{Y \rightarrow X}(m,l)  &=&  H(X_{t_{m+1}}| X_{t_1}\cap \cdots \cap
X_{t_{m}}) - H(X_{t_{m+1}}| X_{t_1}\cap \cdots \cap X_{t_{m}}\cap Y_{t_{m-l+1}}
\cap \cdots \cap Y_{t_{m}} )\nonumber \\[1mm] &=& I(X_{t_{m+1}}; X_{t_1}\cap \cdots \cap
X_{t_{m}}\cap Y_{t_{m-l+1}}\cap \cdots \cap Y_{t_{m}}) - I(X_{t_{m+1}};X_{t_1}
\cap \cdots \cap X_{t_{m}} )\, .\nonumber \\ &&\mbox{\hspace{-15mm}}
\label{III.A.23a}
\end{eqnarray}
The last line of (\ref{III.A.23a}) indicates that $T_{Y \rightarrow X}(m,l)$ represents the following.

\small \begin{tabular}{l l}
  $+$~ & gain of information about $X_{t_{m+1}}$ caused by the
whole history of $X$ and $Y$ up to time $t_m$ \\
  $-$~ & gain of information about $X_{t_{m+1}}$ caused by the
whole history of $X$  up to time $t_m$ \\
  $=$~ & gain of information about $X_{t_{m+1}}$ caused purely by the
whole history of $Y$  up to time $t_m$.
\end{tabular}
\\
\normalsize

\noindent Note that one may equivalently rewrite (\ref{III.A.23a})  as the conditional mutual
information
\begin{eqnarray}
T_{Y \rightarrow X}(m,l)  \ = \ I(X_{t_{m+1}};
Y_{t_{m-l+1}}\cap \cdots \cap Y_{t_{m}}| X_{t_1}\cap \cdots \cap X_{t_{m}})\, .
\label{III.A.23b}
\end{eqnarray}
This shows once more the essence of Schreiber's transfer
entropy, namely, that it describes the gain in information about
$X_{t_{m+1}}$ caused by the whole history of $Y$ (up to time $t_m$)
under the assumption that the whole history of
$X$ (up to time $t_m$) is known. According to the definition of the
conditional mutual information, we can explicitly rewrite Eq.~(\ref{III.A.23b})
as
\begin{eqnarray}
\mbox{\hspace{-10mm}}&&T_{Y \rightarrow X}(m,l)\nonumber \\[2mm] \mbox{\hspace{-7mm}}&&=
\ \sum p(x_{t_1}, \ldots, x_{t_{m+1}}, y_{t_{m-l+1}}, \ldots, y_{t_m}) \log_2
\frac{p(x_{t_{m+1}}|x_{t_1}, \ldots, x_{t_m}, y_{t_{m-l+1}}, \ldots, y_{t_m})}{p(x_{t_{m+1}}|x_{t_1},
\ldots, x_{t_m})}\, ,
\label{III.A.23c}
\end{eqnarray}
where $x_t$ and $y_t$ represent the discrete states at time $t$ of $X$ and $Y$, respectively.

In passing, we may observe from the first line of (\ref{III.A.23a}) that $T_{Y \rightarrow X}\geq 0$
(any extra knowledge in conditional entropy lessens the ignorance). In
addition, due to the Shannon--Gibbs inequality (see, e.g., Ref.~\cite{jaynes}), $T_{Y\rightarrow X}= 0$ only when
\begin{eqnarray}
\frac{p(x_{t_{m+1}}|x_{t_1}, \ldots, x_{t_m},
y_{t_{m-l+1}}, \ldots, y_{t_m})}{p(x_{t_{m+1}}|x_{t_1}, \ldots, x_{t_m})} \ = \ 1\, .
\end{eqnarray}
This, however, means that the history of $Y$ up to time $t_m$ has no influence
on the value of $X_{t_{m+1}}$ or, in other words, there is no information flow
from $Y$ to $X$; i.e., the $Y$ and $X$ time series are independent processes. If there is any kind of
information flow, then $T_{Y \rightarrow X} > 0$.  $T_{Y \rightarrow X}$ is clearly
explicitly non-symmetric (directional) since it measures the degree of dependence of $X$ on $Y$
and not vice versa.

\subsection{Effective transfer entropy~\label{sec4b}}

The effective transfer entropy (ETE) was originally introduced by Marchinski {\em et al.}
in Ref.~\cite{marschinski}, and it was further substantiated in
Refs.~\cite{vejmelka,kwon,lungarella}. The ETE, in contrast to the STE, accounts for
the finite size of a real data set.

In the previous section, we have defined $T_{Y \rightarrow X}(m,l)$ with the history indices $m$ and $l$. In order to view $T_{Y \rightarrow X}$ as a genuine transfer entropy, one should really include in (\ref{III.A.23b}) the whole history of $Y$ and $X$ up to time $t_m$ (i.e., all historical data that may be responsible for cross-correlations with $X_{t_{m+1}}$). The history is finite only if $X$ or/and $Y$ processes are Markovian. In particular, if $X$ is a Markov process of order $m+1$ and and $Y$ is of order $l$, them $T_{Y \rightarrow X}(m,l)$ is a true transfer entropy. Unfortunately most dynamical
systems cannot be mapped to Markovian processes with finite-time memory. For such systems one should take limits $m \rightarrow \infty $ and $l \rightarrow \infty$. In practice, however, the finite size of any real data set hinders this limiting procedure. In order to avoid unwanted finite-size effects, Marchinski proposed the quantity
\begin{eqnarray}
T^{\rm eff}_{Y\rightarrow X}(m,l) \ \equiv \ T_{Y\rightarrow X}(m,l) \ - \ T_{Y_{\rm
schuffled}\rightarrow X}(m,l)\, ,
\label{III.B.28a}
\end{eqnarray}
where $Y_{\rm schuffled}$ indicates the data shuffling via the {\em surrogate data} technique~\cite{surrogate}.  The surrogate data sequence has the same mean, the same variance, the same autocorrelation function, and therefore the same power spectrum as the original sequence, but (nonlinear) phase relations are destroyed. In effect, all the potential correlations between time series $X$ and $Y$ are removed, which means that $T_{Y_{\rm schuffled}\rightarrow X}(m,l)$
should be zero. In practice, this shows itself not to be the case, despite the fact at there is no obvious structure in the data. The non-zero value of $T_{Y_{\rm schuffled}\rightarrow X}(m,l)$ must then be  a byproduct of the finite data set. Definition (\ref{III.B.28a}) then ensures that spurious effects caused by finite $m$ and $l$ are removed.

\section{R\'{e}nyian  transfer entropies~\label{sec7}}


There are various ways in which one can sensibly define a transfer entropy with R\'{e}nyi's information measure $S_q^{(R)}$. The most natural definition is the one based on a $q$-analog of Eqs.~(\ref{III.A.23a})-(\ref{III.A.23b}), i.e.,
\begin{eqnarray}
T_{q;Y \rightarrow X}^{(R)}(m,l)  &=&  S_q^{(R)}(X_{t_{m+1}}| X_{t_1}\cap \cdots \cap X_{t_{m}}) -
S_q^{(R)}(X_{t_{m+1}}| X_{t_1}\cap \cdots \cap X_{t_{m}}\cap Y_{t_1}\cap \cdots \cap Y_{t_{l}} )
\nonumber \\[1mm]
&=& \ I_q^{(R)}(X_{t_{m+1}};  Y_{t_1}\cap \cdots \cap Y_{t_{l}}|
X_{t_1}\cap \cdots \cap X_{t_{m}})\, .
\label{III.B.29aa}
\end{eqnarray}
With the help of (\ref{II.20a}) and (\ref{eq:condMI}) this can be written in an explicit form as
\begin{eqnarray}
&&T_{q;Y \rightarrow X}^{(R)}(m,l)\nonumber \\[2mm] &&= \ \frac{1}{1-q}\log_2 \frac{\sum \varrho_q(x_{t_1},
\ldots, x_{t_m}) p^{\,q}(x_{t_{m+1}}| x_{t_1},\ldots,
x_{t_m} )}{\sum \varrho_q(x_{t_1}, \ldots, x_{t_m}, y_{t_{m-l+1}},
\ldots, y_{t_m}) p^{\,q}(x_{t_{m+1}}|x_{t_1}, \ldots, x_{t_m}, y_{t_{m-l+1}},
\ldots, y_{t_m})}\nonumber
\\[2mm] &&= \ \frac{1}{1-q}\log_2\frac{\sum
\varrho_q(x_{t_1}, \ldots, x_{t_m}) p^{\,q}(y_{t_{m-l+1}}, \ldots, y_{t_m}|x_{t_1},\ldots, x_{t_m})
}{ \sum  \varrho_q(x_{t_1},
\ldots, x_{t_{m+1}})p^{\,q}(y_{t_{m-l+1}}, \ldots, y_{t_m}|x_{t_1}, \ldots, x_{t_{m+1}}) } \, .
\label{III.B.29a}
\end{eqnarray}
Here, $\varrho_q$ is the escort distribution
(\ref{II.15.a}). One can again easily check that in the limit
$q\rightarrow1$ we regain the Shanonnian transfer entropy (\ref{III.A.23c}).

The representation (\ref{III.B.29a}) deserves a few comments. First,
when the history of $Y$ up to time $t_m$ has no influence on the next-time-tick value of $X$ (i.e., on $X_{t_{m+1}}$), then from the
first line in (\ref{III.B.29a}) it follows that $T_{q;Y \rightarrow X}^{(R)}(m,l) =0$, which indicates that no information flows from $Y$ to $X$, as should be expected. In addition, $T_{q;Y \rightarrow X}^{(R)}$
as defined by (\ref{III.B.29aa}) and (\ref{III.B.29a}) takes into
account only the effect of time series $Y$ (up to time $t_m$), while the compound effect of the time series
$X$ (up to time $t_m$) is subtracted (though indirectly present via
correlations that exist between time series $X$ and $Y$).
In the spirit of Section~\ref{sec2c} one may interpret the transfer entropy  $T_{q;Y \rightarrow X}^{(R)}$
as a {\em rating factor} which quantifies
a gain/loss in the risk concerning the behavior of $X$ at the future time $t_{m+1}$  after we take into account
the historical values of a time series $Y$ until $t_{m}$.

Unlike in Shannon's case, $T_{q;Y \rightarrow X}^{(R)} =0$ does not imply independence of the $X$ and $Y$ processes. This is because  $T_{q;Y \rightarrow X}^{(R)}(m,l)$ can also be {\em negative} on account of nonlinear pricing. Negativity of  $T_{q;Y \rightarrow X}^{(R)}$ then simply means that the knowledge of historical values of both $X$ and $Y$ broadens the tail part of the anticipated PDF for the price value $X_{t_{m+1}}$
more than historical values of $X$ only would do. In other words, an extra knowledge of historical values of $Y$ reveals a greater risk in the next time step of $X$ than one would anticipate by  knowing merely the historical data of $X$ alone.


Note that, with our definition (\ref{III.B.29aa}),  $T_{q;Y \rightarrow X}$ is again explicitly directional since it measures the degree of dependence of $X$ on $Y$ and not the
other way around, though in this case we should indicate by an arrow whether the original risk rate about $X_{t_{m+1}}$ was increased or reduced by observing the historical values
of $Y$.

At this stage, one may introduce the effective R\'{e}nyi transfer entropy (ERTE) by following
the same logic as in the Shannonian case. In particular, one can again use the surrogate data technique to define the ERTE as
\begin{eqnarray}
T_{q;Y \rightarrow X}^{(R, \rm{eff} )}(m,l) \ \equiv \ \ T_{q;Y\rightarrow
X}^{(R)}(m,l) \ - \ T_{q;Y_{\rm schuffled}\rightarrow X}^{(R)}(m,l)\, .
\label{III.B.37b}
\end{eqnarray}
Similarly to the RTE,  $T_{q;Y\rightarrow X}^{(R)}(m,l)$ also
accentuates for $q \in (0,1)$ the flow of information that exists between the tail parts of distributions; i.e.,
it describes how marginal events in the time series $Y$  influence
marginal events in the time series $X$. Since most of historical data belong to the central parts of distributions
(typically with well-behaved Gaussian increments), one can
reasonably expect that for $q\in (0,1)$ the transfer entropy $T_{q;Y \rightarrow X}^{(R, \rm{eff} )}(m,l)
\cong T_{q;Y\rightarrow X}^{(R)}(m,l)$, and the surrogate data technique
is not needed. This fact is indeed confirmed in our data analysis presented in the following section.

\section{Presentation of the analyzed data~\label{sec10}}


In the subsequent analysis, we use two types of data set to illustrate the utility of R\'{e}nyi's
transfer entropy. The first data set consists of 11 stock exchange indices, sampled
at a daily (end of trading day) rate. The data set was obtained from Yahoo financial
portal (historical data) with help of the R-code program~\cite{rproject} for the period of time
between 2 January 1998 and 31 December 2009. These data will be used to
demonstrate quantitatively the statistical coherence of all the mentioned indices in the form
of {\em heat maps} and {\em net flows}.

Because we also wish to illustrate our approach quantitatively, we use as a second data set
time series of 183.308 simultaneously recorded data points from two market indices,
namely from the DAX index  and the S\&P500 insex, gathered on a minute-tick basis in the period from
2 April 2008 to 11 September 2009. In our analysis, we use complete records,
i.e., minute data where only valid values for both the DAX index and the S\&P500 index, are admitted:
periods without trading activity (weekends, nighttime, holidays) in one or both stock exchanges
were excluded. This procedure has the obvious disadvantage that records substantially separated
in a real time may become close neighbors in the newly defined time series.
Fortunately, relatively the small number of such ``critical" points compared to the regular ones
prevents a statistically significant error. In addition, due to computer data
massification one may reasonably expect that the trading activity responds almost immediately
to external stimuli. For this reason we have synchronized the data series to an
identical reference time, a master clock, which we take to be Central European Time (CET).

\subsection{Numerical calculation of transfer entropies~\label{sec10.2}}

In order to find the PDF involved in definitions (\ref{III.A.23c}) (respectively, (\ref{III.B.28a})) and (\ref{III.B.29a})
(respectively, (\ref{III.B.37b})) we use the relative-frequency estimate.
For this purpose we divide the amplitude (i.e., stock index) axis into $N$ discrete amplitude bins and
assign to every bin a sample data value. The number of data points per bin
divided by the length of the time series then constitutes the relative frequency which represents the underlying
empirical distribution. In order to implement the R-code in the ETE
and ERTE calculations we partition the data into disjoint equidistant time intervals (blocks), which serve as
a coarse-graining grid. The number of data points we employ in our
calculations is constant in each block. In each block only the arithmetic mean price
is considered in block-dependent computations.

It is clear that the actual calculations depend on the number of  bins chosen (this is also known as the alphabet length).
In Ref.~\cite{marschinski}, it was argued that, in large data sets
such as our time series, the use of alphabets with more than a few symbols is not compatible with the amount
of data at one's disposal. In order to make a connection with existing
results (see Refs.~\cite{kwon2,marschinski}), we conduct calculations at fixed alphabet length $N=3$.

 For a given partition, i.e., fixed $l$, $T_{Y\rightarrow X}(m,l)$ is  a function of the block length $m$.
The parameter $m$ is to be chosen
as large as possible in order to find a stable (i.e., large $m$ independent) value for
$ T_{Y\rightarrow X}(m,l)$; however, due to the finite size of the real time series $X$, it is
required to find a reasonable compromise between unwanted finite sample effects and a
high value for $m$. This is achieved by substituting  $T_{Y\rightarrow X}(m,l)$  with the
effective transfer entropy.
Surrogate data that are needed in definitions  of the ETE (\ref{III.B.28a}) and the ERTE
(\ref{III.B.37b}) are obtained by means of standard R routines~\cite{rproject}.
The effective R\'{e}nyi and  Shannon transfer entropies themselves are explicitly calculated and
visualized with the help of the open-source statistical framework R and its related
R packages for graphical presentations. The calculations themselves are also coded in the
R language.


\subsection{Analyzing the daily data ---
heat maps vs. net information flows~\label{sec10.3}}

The effective transfer entropies $T_{Y \rightarrow X}^{\rm eff}$ and $T_{q;Y \rightarrow X}^{(R, {\rm eff})}$
are calculated between 11 major  stock indices (see the list in
Appendix~A). The results are collected in three tables in Appendix~B and applied in the constructions of
{\em heat maps} and {\em net information flows} in
Figs.~\ref{fig30}--\ref{fig4hb}. In particular, Shannon's information flow  is employed in
Figs.~\ref{fig3b} and \ref{fig4ha}, while R\'{e}nyi's transfer entropy is used in
construction of Figs.~\ref{fig3d}--\ref{fig4hb}. The histogram-based heat map in Fig.~\ref{fig30}
represents the overall run of the 11 aforementioned indices after the filtering procedure. We
have used the RColorBrewer package~\cite{RColorBrewer} from the R statistical environment which employs a
color-spectrum visualization for asset prices. In this case the color runs from the green,
for higher prices, to dark purple, for low price values.

The heat map in Fig.~\ref{fig3b} shows discloses that among the 11 selected markets a substantial
amount of information flows between the Asia--Pacific region (APR) and the US.  One can also
clearly recognize the strong information exchanges between the APR and
European markets and the subdominant information flow between the US and Europe. There is comparably less information
flowing among European
markets themselves.  This can be credited to the fact that the internal European market is typically
liquid and well equilibrated; similarly, a system in thermal equilibrium (far from critical points)
has very little information flow among various parts. An analogous pattern (safe for the NY index) can also be observed
among the US markets.  In contrast, the markets within the APR  mutually exchange a relatively
large volume of information. This might be attributed to a lower liquidity and consequently less balanced
internal APR market.

The heat maps in Figs.~\ref{fig3d} and \ref{fig3dd} bring further understanding. Notably,
we can see that the information flow within APR markets is significantly more
imbalanced between wings of the asset distributions (larger color fluctuations) than
between the corresponding central parts. This suggests low liquidity risks.
A similar though subordinate imbalance in the information transfer
can also be observed  between  the US and APR markets.

%

Understandably more revealing are the net information flows presented in Figs.~\ref{fig4ha},
\ref{fig4h} and \ref{fig4hb}. The net flow $F_{Y\leftrightarrow X}$  is defined as
$F_{Y\leftrightarrow X} \equiv T_{Y\rightarrow X} - T_{X \rightarrow Y}$. This allows
one to visualize more transparently the disparity between the $Y\! \rightarrow \! X$ and
$X\!\rightarrow \! Y$ flows. For instance, in Fig.~\ref{fig4ha} we see that substantially
more information flows from the APR to the US and Europe than vice versa.
Figs.~\ref{fig4h} and \ref{fig4hb} then demonstrate more specifically that the
APR $\rightarrow$ Europe flow is evenly distributed between the central
and  tail distribution parts.  From the net flow in Figs.~\ref{fig4ha},
\ref{fig4h} and \ref{fig4hb} we can also observe an important yet comparably weaker surplus of information
flow from Europe towards the US. This interesting fact  will be further addressed in the following subsection.

Note also that $T_{1.5;SP\&500\rightarrow NY}^{(R)}$,
$T_{0.8;SP\&500\rightarrow NY}^{(R)}$ and $T_{0.8;NY\rightarrow DJ}^{(R)}$
have negative values. These exceptional behaviors can be partly
attributed to the fact that both the SP\&500 and DJ indices are built from
indices that are also present in the NY index and hence one might expect unusually strong
coherence between these indices.
From Section~\ref{sec7} we know that negative values of  the ERTE  imply
a higher risk involved in a next-time-step asset-price behavior
than could be predicted (or expected) without knowing the historical values of the source time series.
The observed negativity of $T_{0.8;X\rightarrow Y}^{(R)}$ thus
means that when some of the ignorance is elevated by observing the time series $X$
a higher risk reveals itself in the nearest-future behavior of the asset price $Y$.
Analogously, negativity of $T_{1.5;X\rightarrow Y}^{(R)}$ corresponds to a risk enhancement of
the non-risky (i.e. close-to-peak) part of the underlying PDF.



\subsection{Minute-price information flows~\label{sec10.d}}

Here we analyze the minute-tick historic records of the DAX and S\&P500 indices  collected
over the period of 18 months from 2 April 2008 to 11 September 2009. The coarse-grained overall run of both indices after the filtering procedure is depicted in the histogram-based heat map in Fig.~\ref{fig30}.

Without any a prior knowledge about the Markovian (or non-Markovian) nature of the data series, we consider the order of the Markov process for both the DAX and S\&P500 stocks to be identical, i.e., the price memory of both indices is considered to be the same. The latter may be viewed as a ``maximally unbiased'' assumption. At this stage we eliminate the surrogate data and consider the RTE alone. The corresponding RTEs for $q=1.5$ and $q=0.8$ as functions of block lengths are shown in Figs.~\ref{fig4d} and \ref{fig4dd}, respectively. There we can clearly recognize that for $m\sim 200-300$ minutes there are no new correlations between the DAX and S\&P500 indices. So, the underlying Markov process has order (or memory) roughly $200-300$ minutes.

The aforementioned result is quite surprising in view of the fact that autocorrelation
functions of stock market returns typically decay exponentially with a characteristic time
of the order of minutes (e.g., $\sim$ 4 mins for the S\&P500~\cite{mategna,jizba4}), so the returns are basically uncorrelated random variables. Our result, however, indicates that two markets can be intertwined for much longer. This situation is actually not so surprising when we realize that empirical analysis of financial data asserts (see, e.g.,~\cite{Liu}) that autocorrelation functions of higher-order correlations for asset returns have longer decorrelation time which might span up to years (e.g., a few months in the case of volatility for the S\&P500~\cite{jizba4}). It is indeed a key advantage of our approach that the {\em nonlinear} nature of the RTE naturally
allows one to identify the existing long-time cross-correlations between financial markets.

In Fig.~\ref{fig4c}, we depict the empirical dependence of the ERTE on the parameter $q$. Despite the fact that the RE itself is  a monotonically decreasing function of $q$ (see, e.g., Ref.~\cite{renyi0}) this is generally not the case for the ERTE (nor for the conditional RE).
Indeed, the ERTE represents a difference of two REs with an identical $q$ (see Eq.~(\ref{III.B.29aa})), and as such it may be neither monotonic nor decreasing. The
functional dependence of the ERTE on $q$ nevertheless serves as an important indicator of how quickly REs involved  change with $q$.



The results reproduced in Fig.~\ref{fig4c} quantitatively confirm the expected asymmetry in the
information flow between the US and European markets. However, since the US
contributes more than half of the world's trading volume, it could be anticipated that there
is a stronger information flow from big US markets towards both European and APR
markets. Yet, despite the strong US trading record, our ERTE approach indicates that the
situation is not so straightforward when the entropy-based information flow is considered as a
measure of market cross-correlation. Indeed, from  Figs.~\ref{fig4ha}, \ref{fig4h} and \ref{fig4hb} we could observe that there is a noticeably stronger information flow from the European and APR markets to the U.S. markets than than vice versa. Fig.~\ref{fig4c} extends the validity of this observation also to short time scales of the order of minutes. In particular, from Fig.~\ref{fig4c} we clearly see that flow from the DAX to the S\&P500 is stronger than the reverse flow. It is also worth of noting that this Europe--US flow is positive for all values of $q$, i.e., for all distribution sectors, with a small bias towards tail parts of the underlying distribution.

\section{Concluding remarks~\label{SEc12}}

Transfer entropies have been repeatedly utilized in the quantification of statistical coherence
between various time series with prominent applications in financial markets. In contrast
to previous works in which transfer entropies have been exclusively considered only in the
context of Shannon's information theory, we have advanced here the notion of R\'{e}nyi's (i.e.
non-Shannonian) transfer entropy. The latter is defined in a close analogy with Shannon's
case, i.e., as the information flow (in bits) from $Y$ to $X$ ignoring static correlations due
to the common historical factors such as external agents or forces. However, unlike Shannon's
transfer entropy, where the information flow between two (generally cross-correlated)
stochastic processes takes into account the whole underlying empirical price distribution, the
RTE describes the information flow only between certain pre-decided parts of two
price distributions involved. The distribution sectors in question can be chosen when R\'{e}nyi's parameter $q$ is set in accordance with Campbell's pricing theorem. Throughout this paper we have demonstrated that the RTE thus defined has many specific properties that are desirable for
the quantification of an information flow between two interrelated stochastic systems. In particular, we have shown that the RTE can serve as an efficient rating factor which quantifies
a gain or loss in the risk that is inherent in the passage from $X_{t_m}$ to $X_{t_{m+1}}$ when a new information, namely historical values of a time series $Y$ until time $t_m$, is taken into account. This
gain/loss is parameterized by a single parameter, the R\'{e}nyi $q$ parameter, which serves as a ``zooming index" that zooms (or emphasizes) different sectors of the underlying empirical
PDF. In this way one can scan various sectors of the price distribution and analyze associated
information flows. In particular, the fact that one may separately scrutinize information
fluxes between tails or central-peak parts of asset price distributions  simply by setting $q< 1$ or $q>1$, respectively, can be employed, for example, by financial institutions to quickly analyze the global (across-the-border) information flows and use them to redistribute their risk. For instance, if an American investor observes that a certain market, say the S\&P500, is going down and he/she knows that the corresponding NASDAQ ERTE for $q < 1$ is low, then he/she
does not need to relocate the portfolio containing related assets rapidly, because the influence is in this case slow. Slow portfolio relocation is generally preferable, because fast relocations are always burdened with excessive transaction costs. Let us stress that this type of conduct could not be deduced from Shannon's transfer entropy alone. In fact, the ETE suggests a fast (and thus expensive) portfolio relocation as a best strategy (see Figs.~\ref{fig4ha},\ref{fig4h} and \ref{fig4hb}).

Let us stress that  applications of transfer entropies presented
quantitatively support the observation that more information flows from the Asia--Pacific
region towards the US and Europe than vice versa, and this holds for transfers
between both peak parts and wing parts of asset PDFs; i.e., the US and European markets are
more prone to price shakes in the Asia--Pacific sector than the other way around. Besides,
information-wise  the US market is more influenced by the European one than in reverse.
This interesting observation can be further substantiated by our DAX versus S\&P500 analysis,
in which we have seen that the influx of information from Europe is to a large extend due to a
tail-part transfer. The peak-part transfer is less pronounced. So, although
the US contributes more than half of the world's trading volume, our results
indicate that this is not so with information flow. In fact, the US markets seem to
be prone to reflect a marginal (i.e., risky) behavior in both European and APS markets. Such
a fragility does not seem to be reciprocated.
This point definitely deserves further closer analysis.

Finally, one might be interested in how the RTE presented here compares with other correlation tests.
The usual correlation tests take into account either the lower-order correlations (e.g., time-lagged cross-correlation test and Arnhold {\em et al.} interdependence test) or they try to address the causation issue between  bivariate time series (e.g., Granger causality test or Hacker and Hatemi-J causality test). Since the RTE allows one to compare only certain parts of the underlying distributions it also works implicitly with high-order correlations, and for the same reason it cannot affirmatively answer the causation issue. In many respects such correlation tests bring complementary information with respect to the RTE approach.
More detailed discussion concerning multivariate time series and related correlation tests will be presented elsewhere.



\section*{Acknowledgments}

This work was partially supported by the Ministry of Education of
the Czech Republic (research plan MSM 6840770039), and by the Deutsche
Forschungsgemeinschaft under grant Kl256/47.

\newpage

\appendix \section*{Appendix A\label{ap1}}
\setcounter{section}{1}\setcounter{equation}{0}

In this appendix we provide a brief glossary of the indices used in the main text.
The notation presented here conforms with the notation typically
listed in various on-line
financial portals (e.g., Yahoo financial portal).
\vspace{10mm}

{\small \begin{tabular}{|l|p{12cm}|l|}
\hline {\bf Indices} & {\bf Description} & {\bf Country}
\\
\hline \hline {\bf GSPC} & Standard and Poor 500 (500 stocks actively traded
in the U.S.)& USA
\\
\hline {\bf GDAXI} & Dax Indices (stock of 30 major  German companies) & Germany
\\
\hline {\bf ATX} & The Austrian Traded Index is the most important stock
market index of the Wiener B\"{o}rse. The ATX is a price index and currently consists of 20 stocks. & Austria
\\
\hline {\bf SSMI} & The Swiss Market Index is a
capitalization-weighted index of the 20 largest and most liquid stocks. It represents about 85\%
of the free-float market capitalization of the Swiss equity market.
 & Swiss
 \\
\hline {\bf AORD}  & All Ordinaries represents the 500 largest companies in the Australian
equities market. Index constituents are drawn from eligible  companies listed on the
Australian Stock Exchange & Australia
\\
\hline {\bf BSESN} & The BSE Sensex is a market capitalized index that tracks 30 stocks
from the Bombay Stock Exchange. It is the second
largest exchange of India in terms of volume and first in terms of shares listed.
& India
\\
\hline {\bf HSI} & The Hang Seng Index denoted in Hong Kong stock market. It is used
to record and monitor daily changes of the largest companies of the Hong Kong stock market.
It consist of 45 Companies.  &Hong Kong
\\
\hline {\bf N225} & Nikkei 225  is a stock
market index for the Tokyo Stock Exchange. It is a price-weighted average (the unit is yen),
and the components are reviewed once a year. Currently, the Nikkei is the most widely
quoted average of Japanese equities, similar to the Dow Jones Industrial Average.& Japan
\\
\hline {\bf DJA} &The Dow Jones Industrial Average also referred to as the Industrial
Average, the Dow Jones, the Dow 30, or simply as the Dow; is one of several U.S.
stock market indices. First published in 1887. &USA
\\
\hline {\bf NY} & iShares NYSE 100 Index
is an exchange trading fund, which is a security that tracks a basket of assets,
but trades like a stock. NY tracks the SE U.S. 100; this equity index measures the performance of
the largest 100 companies listed on the New York Stock Exchange (NYSE). & USA
\\
\hline {\bf IXIC} & The Nasdaq Composite is a stock market index of all of the common stocks and
similar securities (e.g., ADRs, tracking stocks, limited partnership interests)
listed on the NASDAQ stock market, it has over 3.000 components. &USA
\\
\hline
\hline
\end{tabular}
\normalsize}

 \section*{Appendix B\label{ap2}}
%
%
In this appendix we specify explicit values of effective transfer entropies that are employed in
Section~\ref{sec10}. These are calculated for alphabet with $N=3$.
\begin{figure}[ht]
\begin{center}
\includegraphics*[width=15cm]{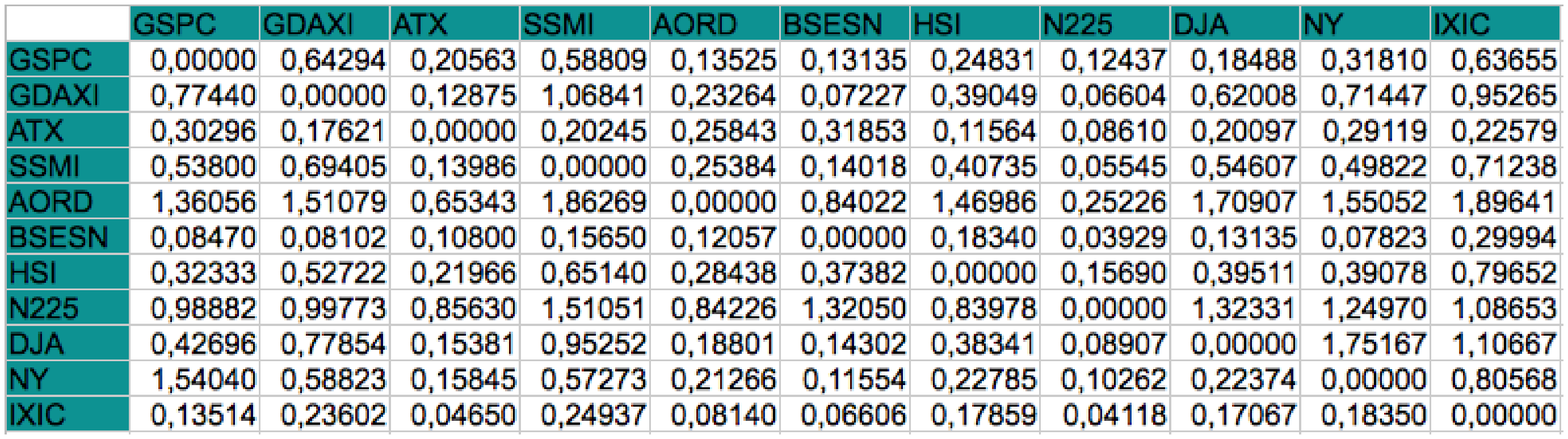}
\end{center} \vspace{-.5cm}
\caption{Numerical data for the ETE that are used to generate Figs.~\ref{fig3b} and \ref{fig4ha}.} \label{fig30b}
\end{figure}
\begin{figure}[ht]
\begin{center}
\includegraphics*[width=15cm]{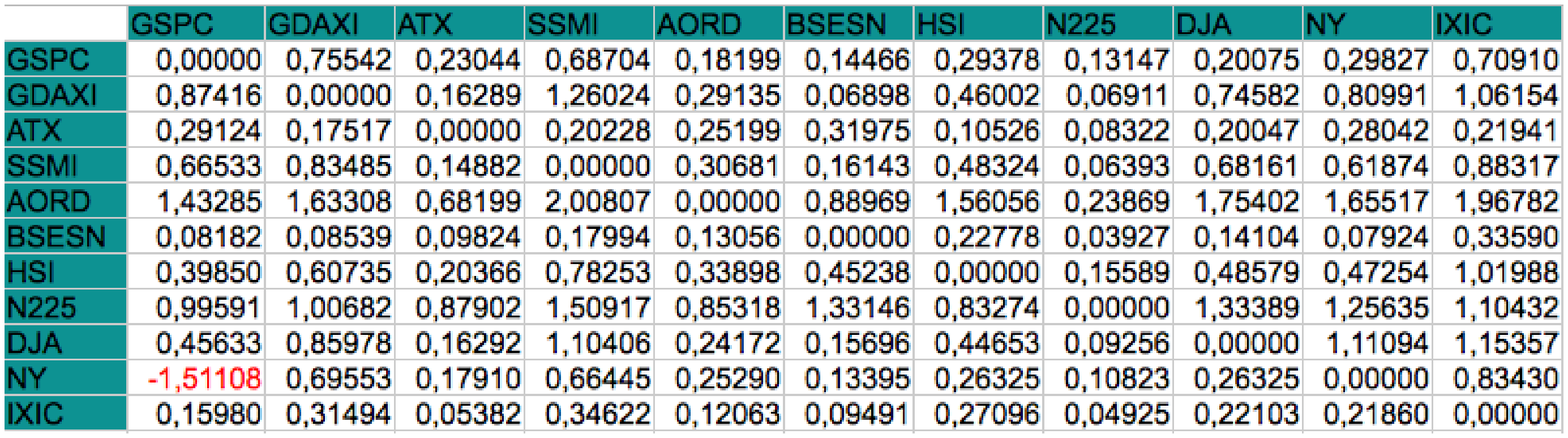}
\end{center} \vspace{-.5cm}
\caption{Numerical data for the ERTE that are used to
generate Figs.~\ref{fig3d} and \ref{fig4h}; $q =1.5$.}
\label{fig30c}
\end{figure}
\begin{figure}[ht]
\begin{center}
\includegraphics*[width=15cm]{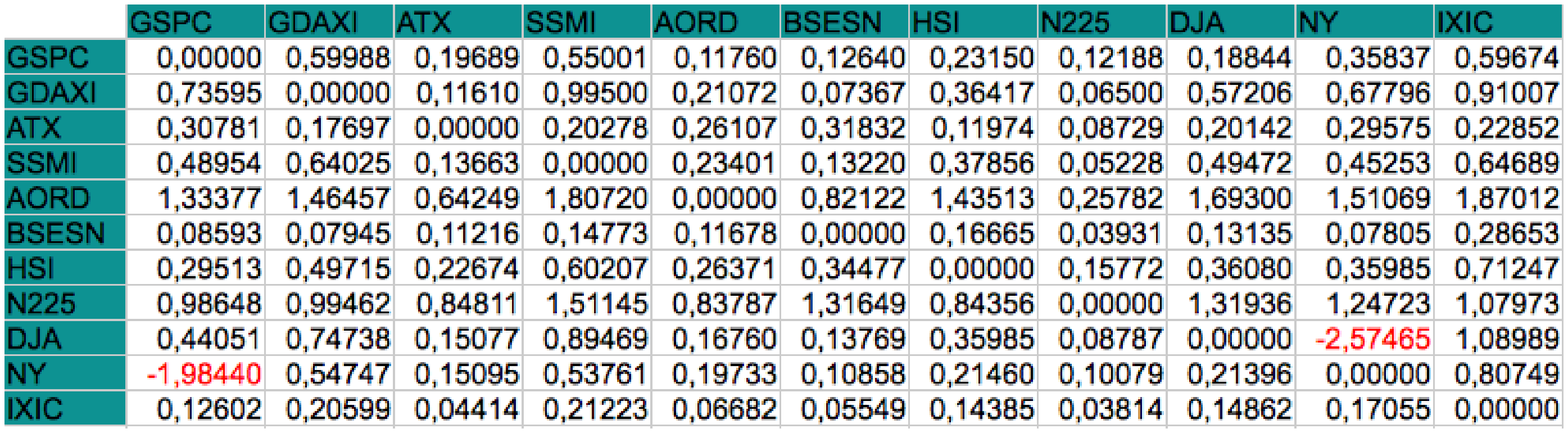}
\end{center} \vspace{-.5cm}
\caption{Numerical data for the ERTE that are used to generate
Figs.~\ref{fig3dd} and~\ref{fig4hb}; $q =0.8$.}
\label{fig30cb}
\end{figure}


%
\newpage

\begin{figure}[ht]
\begin{center}
\scalebox{2.5}[1.75]
{\includegraphics*[width=6.5cm]{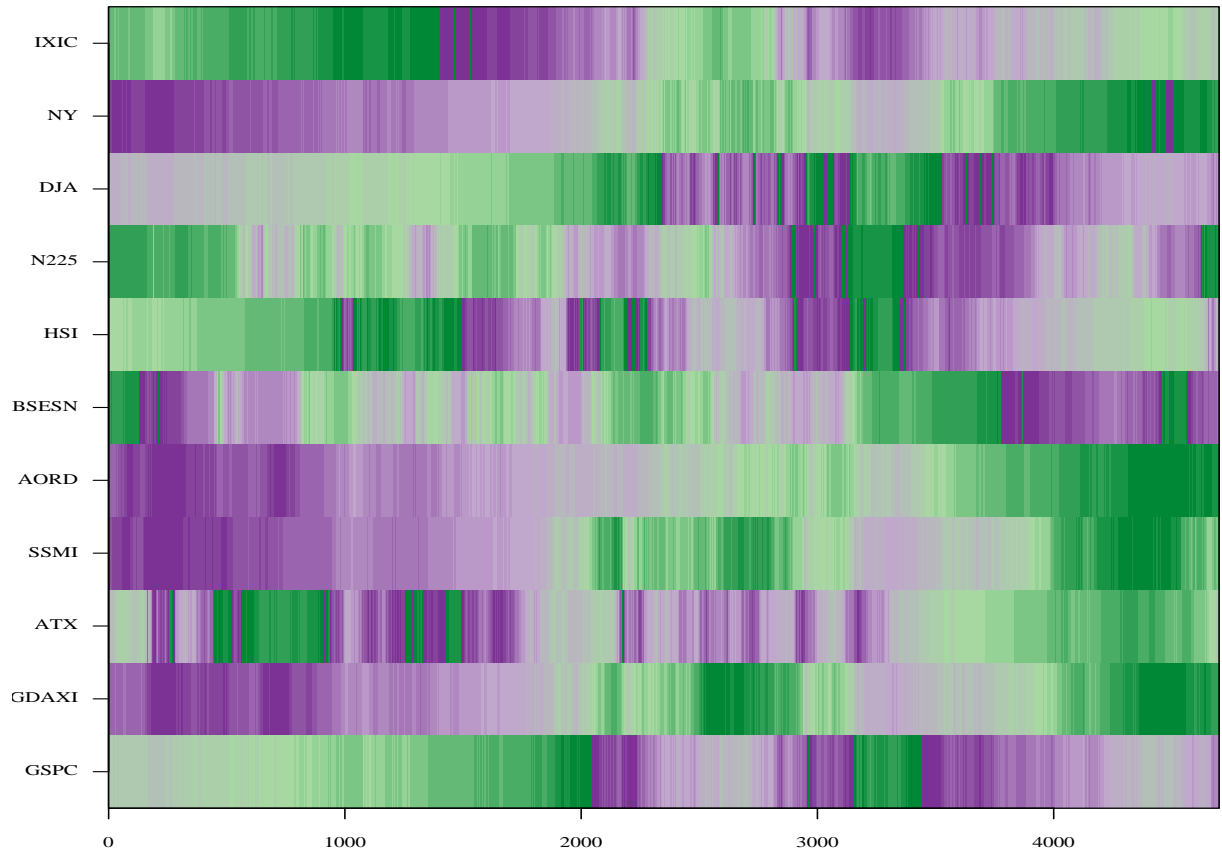}}
\end{center}
\vspace{-.5cm}
\caption{Histogram-based (i.e., non-entropic) heat map between
the 11 stock indices listed in Appendix~A.}
\label{fig30}
\end{figure}

\newpage
\begin{figure}[ht]
\begin{center}
\scalebox{2.5}[1.75]
{\includegraphics*[width=7cm]{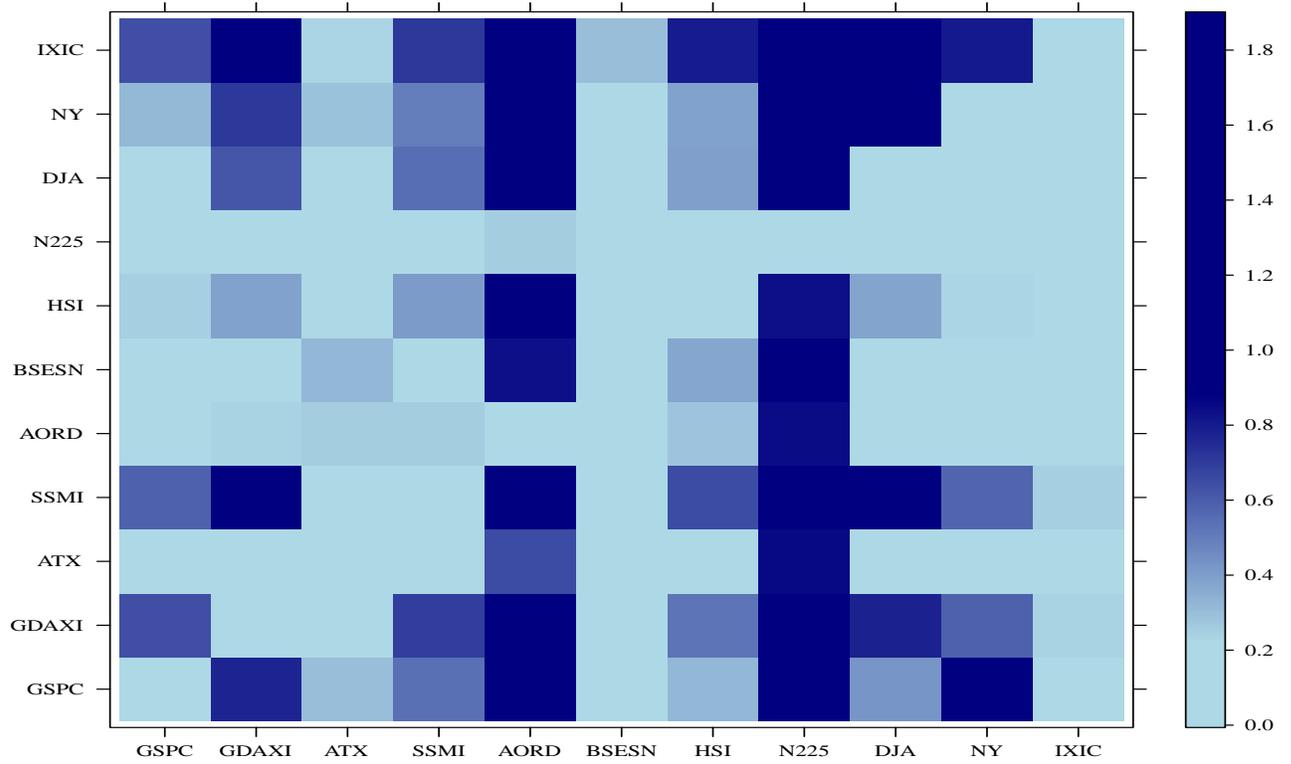}}
\end{center} \vspace{-.5cm}
\caption{Heat map of Shannon's effective entropy between the 11 stock indices
listed in Appendix~A. Alphabet size $N=3$.}
\label{fig3b}
\end{figure}

\newpage

\begin{figure}[ht]
\begin{center}
\includegraphics*[width=17cm]{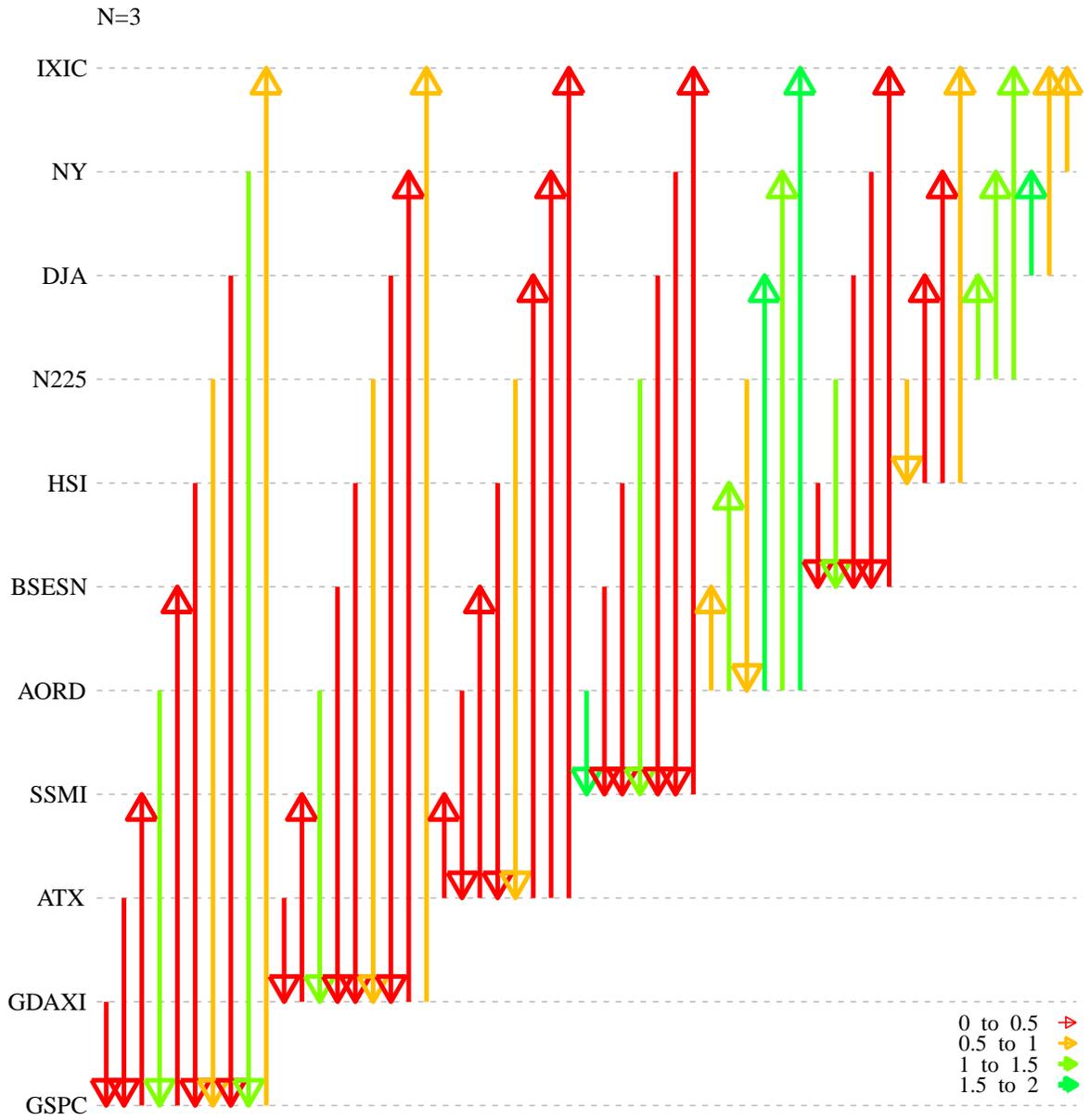}
\end{center}
\vspace{-.5cm}
\caption{Net flow $F_{Y\leftrightarrow X}$ of
effective Shannon transfer entropies between the 11 stock indices listed in Appendix~A. Alphabet size $N=3$.}
\label{fig4ha}
\end{figure}
\newpage

\begin{figure}[ht]
\begin{center}
\scalebox{2.5}[1.75]
{\includegraphics*[width=7cm]{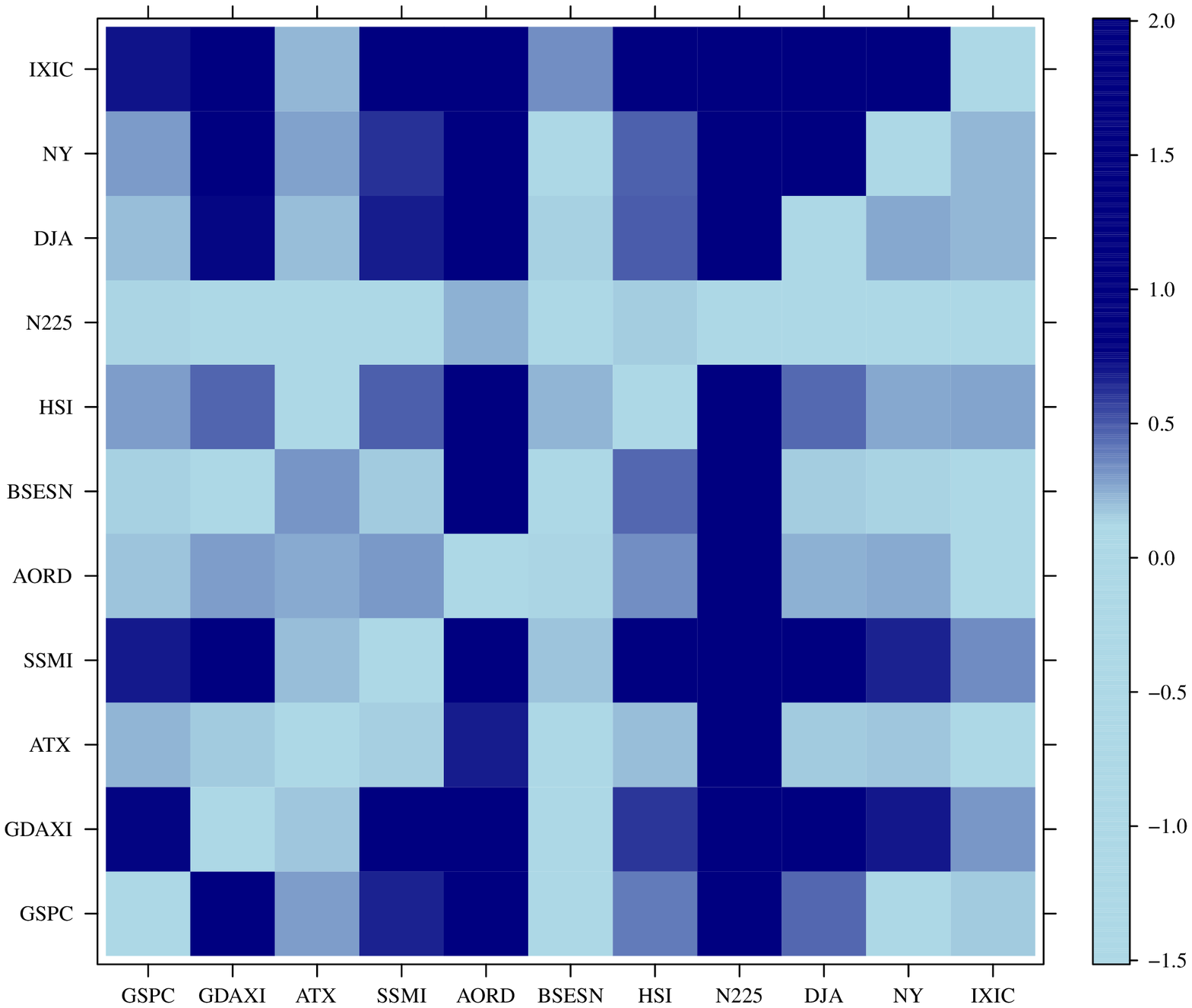}}
\end{center}
\vspace{-.5cm}
\caption{Heat map of  R\'{e}nyi's effective entropy between the 11 stock indices listed in Appendix~A; $q = 1.5$. Alphabet size $N=3$.}
\label{fig3d}
\end{figure}

\newpage

\begin{figure}[ht]
\begin{center}
\includegraphics*[width=17cm]{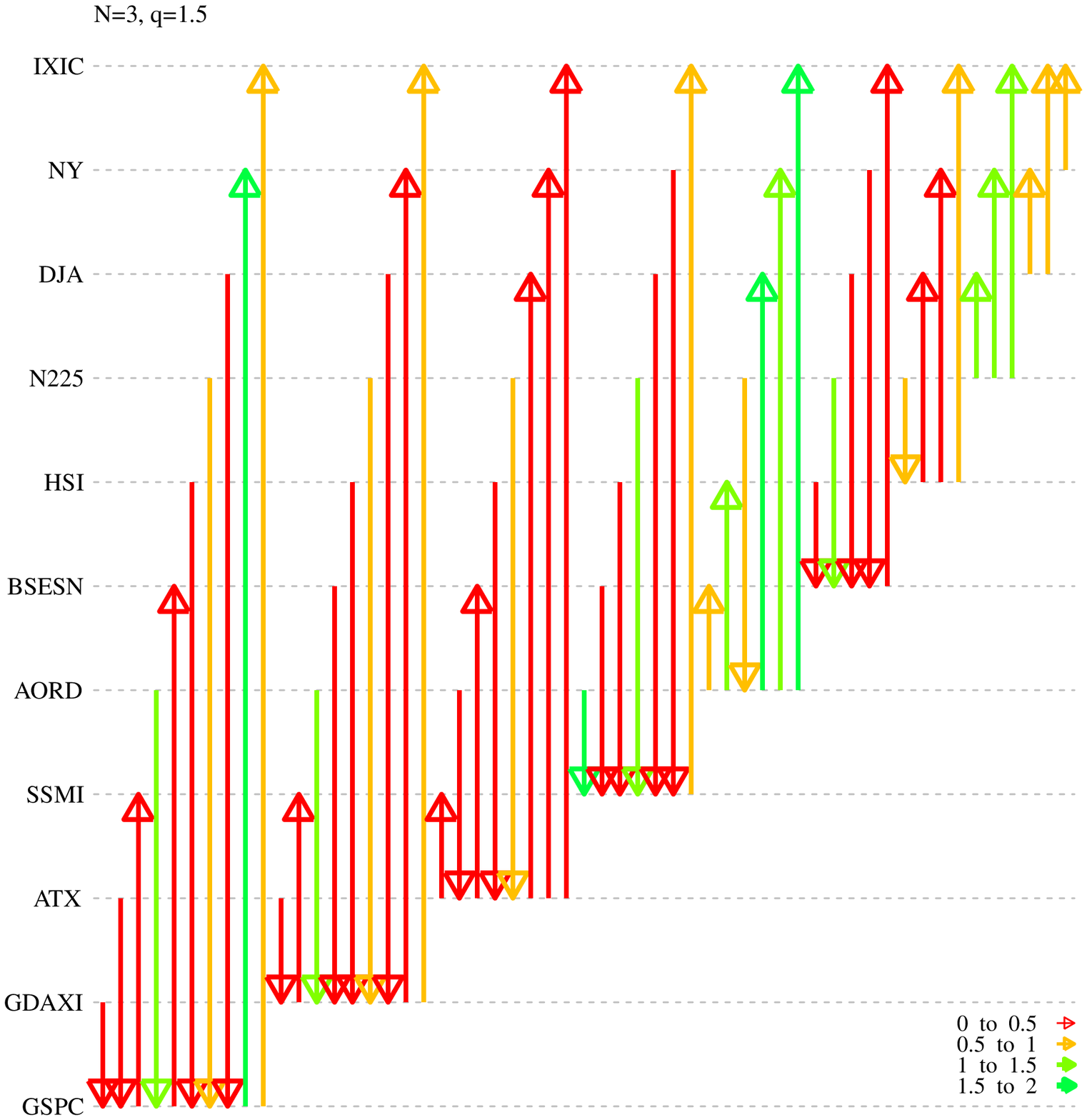}
\end{center}
\vspace{-.5cm}
\caption{Net flow $F_{Y\leftrightarrow X}$ of the effective R\'{e}nyi  transfer entropies listed in Appendix~A; $q=1.5$. Alphabet size $N=3$.}
\label{fig4h}
\end{figure}
\newpage

\begin{figure}[ht]
\begin{center}
\scalebox{2.5}[1.75]
{\includegraphics*[width=7cm]{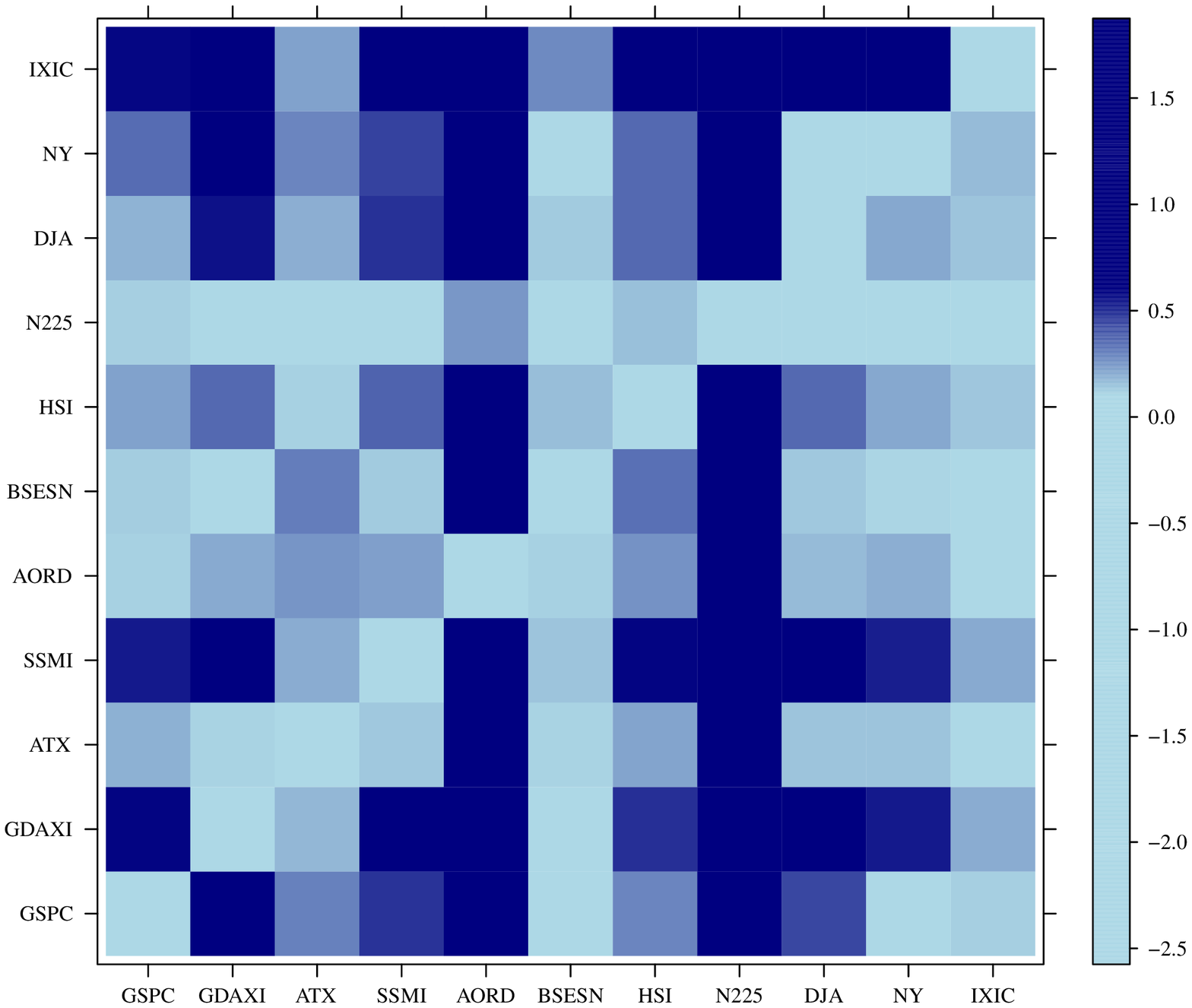}}
\end{center}
\vspace{-.5cm}
\caption{Heat map of  R\'{e}nyi's effective entropy between the 11 stock indices listed in Appendix~A; $q = 0.8$. Alphabet size $N=3$.}
\label{fig3dd}
\end{figure}
\newpage

\begin{figure}[ht]
\begin{center}
\includegraphics*[width=17cm]{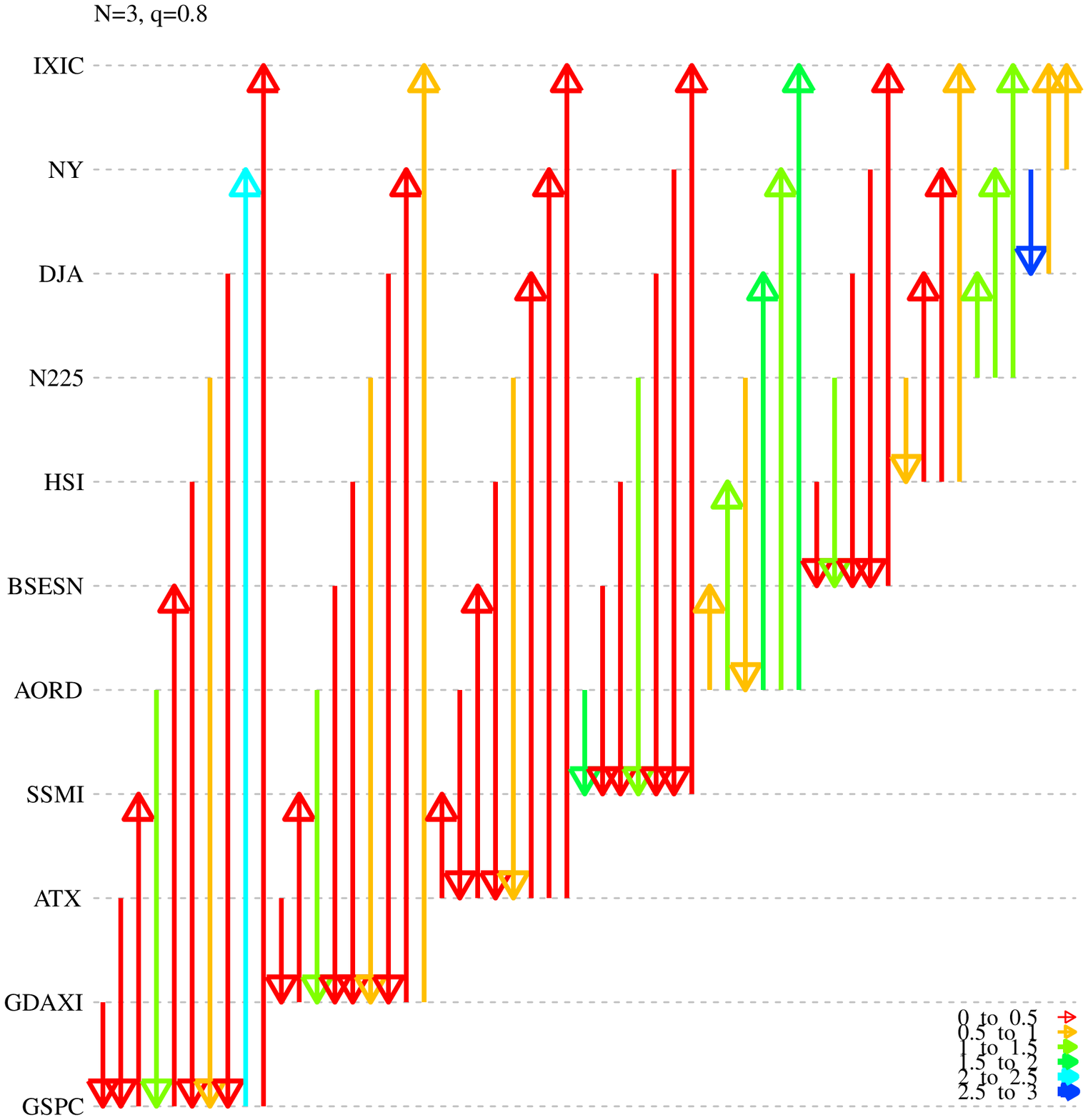}
\end{center} \vspace{-.5cm}
\caption{Net flow $F_{Y\leftrightarrow X}$ of effective
R\'{e}nyi  transfer entropies listed in Appendix~A; $q=0.8$. Alphabet size $N=3$.}
\label{fig4hb}
\end{figure}

\newpage

\begin{figure}[ht]
\begin{center}
\includegraphics*[width=11.3cm]{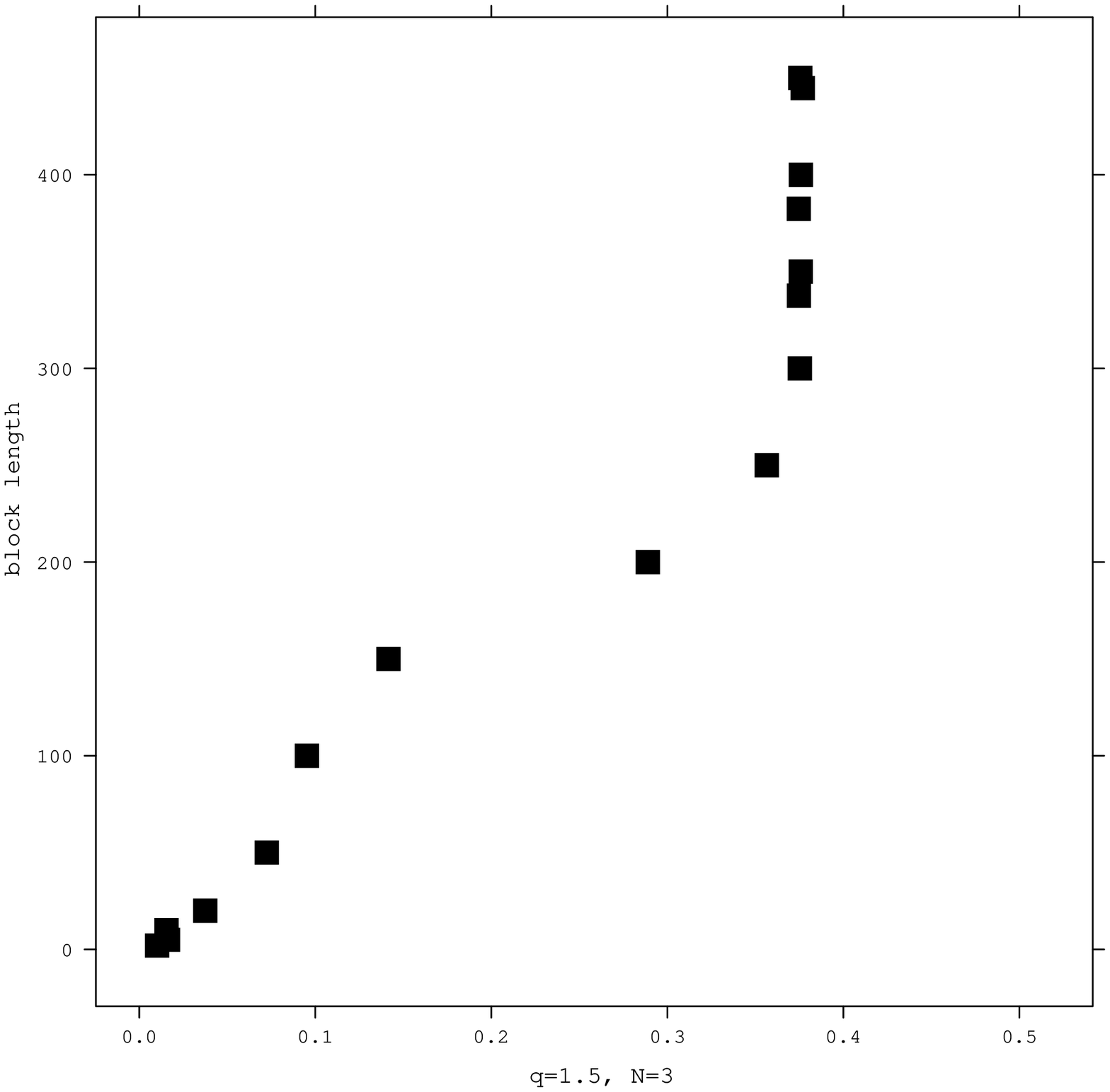}
\end{center}
\vspace{-.3cm}
\caption{Effective R\'{e}nyi transfer entropy $T_{1.5;SP\&500\rightarrow DAX}^{(R)}(m,m)$
for a $3$-letter alphabet as a function of the block length $m$.
DAX and SP\&500 minute prices are employed. The correlation time is between $200-300$ minutes. }
\label{fig4d}
\end{figure}
\begin{figure}[ht]
\begin{center}
\includegraphics*[width=11.3cm]{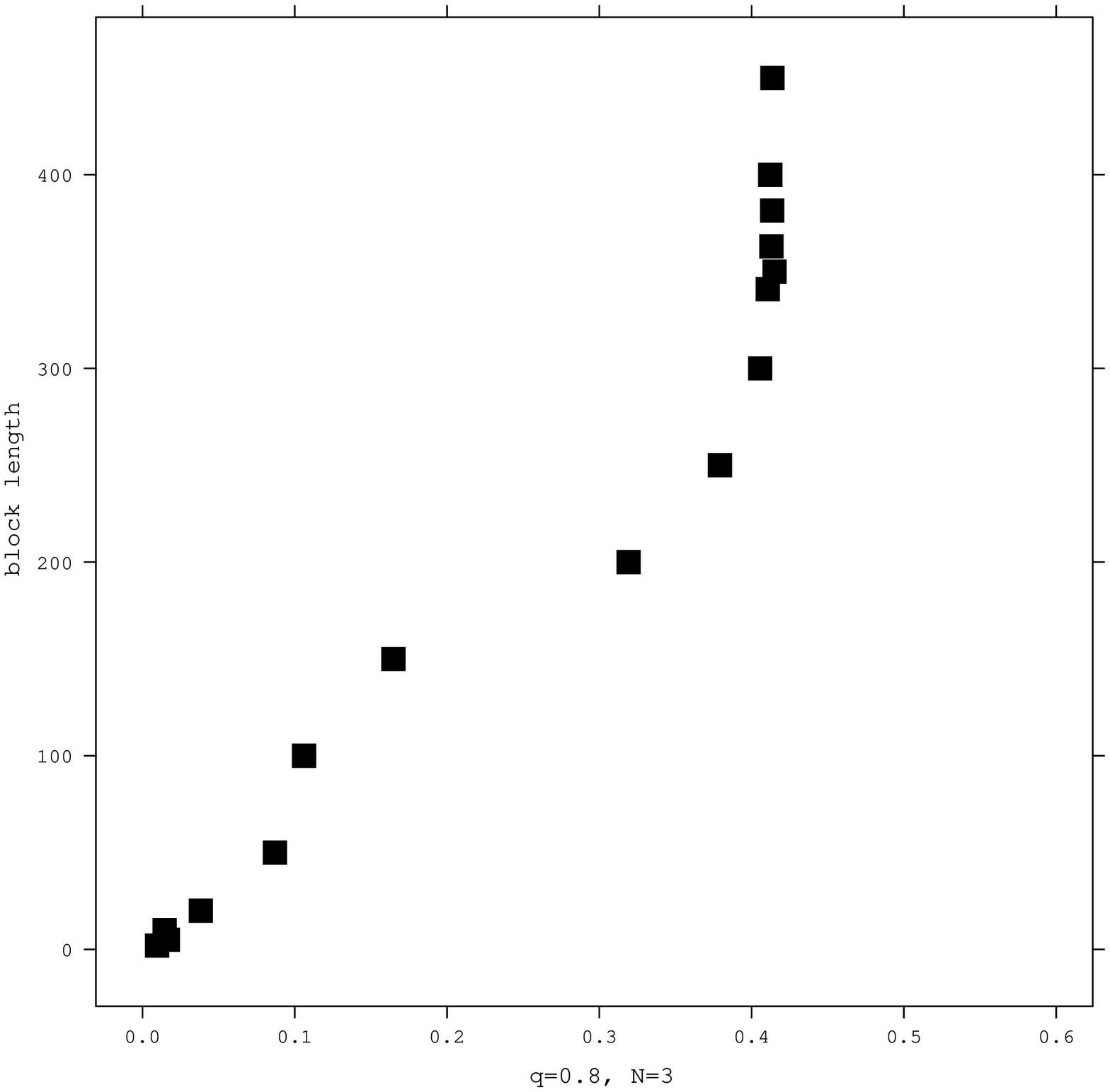}
\end{center}
\vspace{-.3cm}
\caption{Effective R\'{e}nyi transfer entropy $T_{0.8;SP\&500\rightarrow DAX}^{(R)}(m,m)$
for a $3$-letter alphabet as a function of the block length $m$. DAX and SP\&500 minute prices are employed.
The correlation time is between $200-300$ minutes.}
\label{fig4dd}
\end{figure}
%

\newpage 

\begin{figure}[ht]
\begin{center}
\includegraphics*[width=11.5cm]{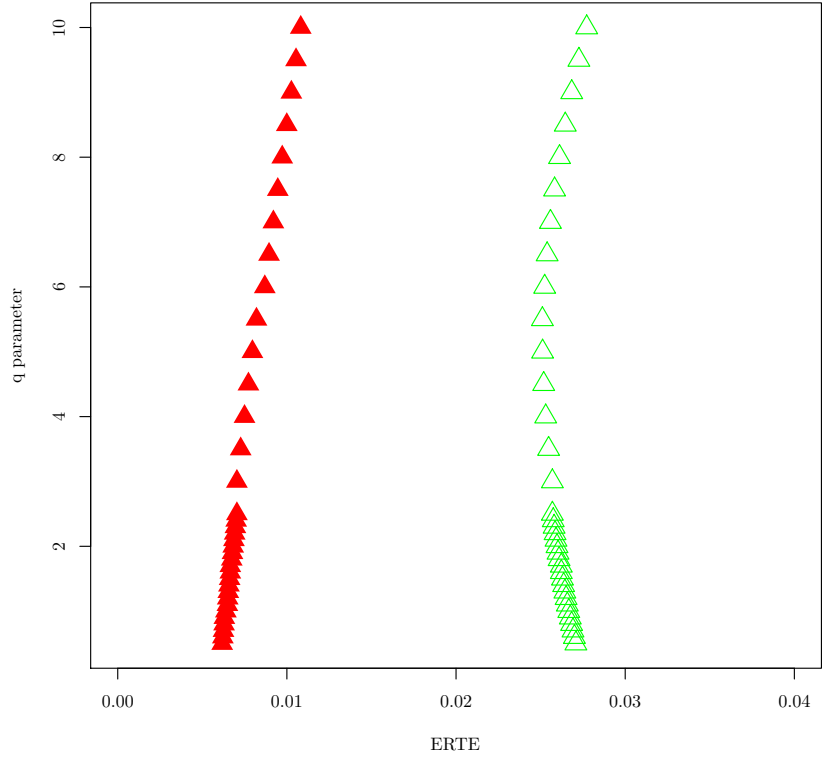}
\end{center}
\vspace{-.5cm}
\caption{The ERTE as a function of $q$. The alphabet size is set to
$N = 3$. DAX and SP\&500 minute prices are employed.
The red curve corresponds to $T_{q;SP\&500\rightarrow DAX}^{(R, \rm{eff} )}(m,m)$
while the green curve denotes $T_{q;DAX\rightarrow SP\&500}^{(R, \rm{eff} )}(m,m)$.}
\label{fig4c}
\end{figure}

%


\begin{thebibliography}{99}
\bibitem{gibbs} Maximal entropy principle was in statistical thermodynamic introduced by
J.W.~Gibbs under the name: ``The fundamental hypothesis of equal a priori probabilities
in the phase space".
\bibitem{expweight} This exponential weighting is also known as a
Kolmogorov--Nagumo averaging. While the linear averaging is given
by $\langle X \rangle = \sum_{x\in X}^W p(x) x$, the exponential
weighting ia defined as $\langle X \rangle_{\rm exp} = \phi^{-1}\left(\sum_{x\in X}^W p(x)
\phi(x) \right)$ with $\phi(x) = 2^{\beta x}$.
The $\beta = (1-q)/q$ factor is known as Campbell exponent.
\bibitem{note} One can map $S_q^{(R)}$ with $q\geq 1$ to $S_q^{(R)}$ with $0 < q \leq1$ via duality
$q\leftrightarrow 1/q$ that exists between $\varrho_q$ and ${\mathcal{P}}$.
In fact, we can observe that $S_{1/q}^{(R)}(\varrho_q)  =  S_q^{(R)}({\mathcal{P}})$
and $S_{q}^{(R)}(\varrho_{1/q})  =  S_{1/q}^{(R)}({\mathcal{P}})$. So $S_q^{(R)}$ with
$q\geq 1$ and $S_q^{(R)}$ with $0 < q \leq 1$ carry equal amount of information.
\end{thebibliography}
\end{document}